\documentclass[epj]{svjour}
\usepackage{graphics}
\usepackage{graphicx}  
\usepackage{epsfig} 
\usepackage{subfigure}
\usepackage{lineno}  
\usepackage{amsmath,amssymb}
\usepackage{comment}
\usepackage{multirow}
\usepackage{dcolumn}
\usepackage{bm}
\usepackage{setspace}
\includecomment{printrobustnotes}
\includecomment{printallnotes}
\usepackage{xspace}
\usepackage{color}
\def\kforty{$\rm ^{40}K$~}

\def\utwothirtyeight{$\rm ^{238}U$~}
\def\thtwothirtytwo{$\rm ^{232}Th$~}
\def\rntwotwotwo{$\rm ^{222}Rn$~}

\def\ionetwentyfive{$\rm ^{125}I$~}

\def\pbtwoten{$\rm ^{210}Pb$~}
\def\potwoten{$\rm ^{210}Po$~}

\def\amtwofortyone{$\rm ^{241}Am$~}

\def\kforty{$\rm ^{40}K$~}

\def\utwothirtyeight{$\rm ^{238}U$~}
\def\thtwothirtytwo{$\rm ^{232}Th$~}
\def\rntwotwotwo{$\rm ^{222}Rn$~}

\def\ionetwentyfive{$\rm ^{125}I$~}

\def\pbtwoten{$\rm ^{210}Pb$~}
\def\potwoten{$\rm ^{210}Po$~}

\def\amtwofortyone{$\rm ^{241}Am$~}

\def\cosixty{$\rm ^{60}Co$~}

\newcommand{\gev}{\ensuremath{{\rm GeV}/{\rm c}^2}}

\newcommand{\kgyr}{\ensuremath{{\rm kg\cdot yr}}}

\begin{document}

\title{Initial Performance of the COSINE-100 Experiment}

\author{
  G.~Adhikari\inst{1} \and
  P.~Adhikari\inst{1} \and
  E.~Barbosa de Souza\inst{2} \and
  N.~Carlin\inst{3} \and
  S.~Choi\inst{4} \and
  W.Q.~Choi\inst{5}\thanks{\emph{Present address:} Institut f\"{u}r Experimentelle Kernphysik, Karlsruher Institut f\"{u}r Technologie (KIT), Eggenstein-Leopoldshafen 76344, Germany} \and
  M.~Djamal\inst{6} \and
  A.C.~Ezeribe\inst{7} \and
  C.~Ha\inst{8}\thanks{Corresponding Author : {changhyon.ha@gmail.com}} \and
  I.S.~Hahn\inst{9} \and
  A.J.F.~Hubbard\inst{2}\thanks{\emph{Present address:} Department of Physics \& Astronomy, Northwestern University, Evanston, IL 60208, USA} \and
  E.J.~Jeon\inst{8} \and
  J.H.~Jo\inst{2} \and
  H.W.~Joo\inst{4} \and
  W.G.~Kang\inst{8} \and
  W.~Kang\inst{10} \and
  M.~Kauer\inst{11} \and
  B.H.~Kim\inst{8} \and
  H.~Kim\inst{8} \and
  H.J.~Kim\inst{12} \and
  K.W.~Kim\inst{8} \and
  M.C.~Kim\inst{10}\thanks{\emph{Present address:} Department of Physics, Chiba University, Chiba 263-8522, Japan} \and
  N.Y.~Kim\inst{8} \and
  S.K.~Kim\inst{4} \and
  Y.D.~Kim\inst{8,1} \and
  Y.H.~Kim\inst{8,13} \and
  V.A.~Kudryavtsev\inst{7} \and
  H.S.~Lee\inst{8} \and
  J.~Lee\inst{8} \and
  J.Y.~Lee\inst{12} \and
  M.H.~Lee\inst{8} \and
  D.S.~Leonard\inst{8} \and
  K.E.~Lim\inst{2} \and
  W.A.~Lynch\inst{7} \and
  R.H.~Maruyama\inst{2} \and
  F.~Mouton\inst{7} \and
  S.L.~Olsen\inst{8} \and
  H.K.~Park\inst{8} \and
  H.S.~Park\inst{13} \and
  J.S.~Park\inst{8}\thanks{\emph{Present address:} High Energy Accelerator Research Organization (KEK), Ibaraki 319-1106, Japan} \and
  K.S.~Park\inst{8} \and
  W.~Pettus\inst{2}\thanks{\emph{Present address:} Center for Experimental Nuclear Physics and Astrophysics and Department of Physics, University of Washington, Seattle, WA 98195, USA} \and
  Z.P.~Pierpoint\inst{2} \and
  H.~Prihtiadi\inst{6} \and
  S.~Ra\inst{8} \and
  F.R.~Rogers\inst{2}\thanks{\emph{Present address:} Department of Physics, Massachusetts Institute of Technology, Cambridge, MA 02139, USA} \and
  C.~Rott\inst{10} \and
  A.~Scarff\inst{7}\thanks{\emph{Present address:} Department of Physics and Astronomy, University of British Columbia, Vancouver, BC V6T 1Z1, Canada} \and
  N.J.C.~Spooner\inst{7} \and
  W.G.~Thompson\inst{2} \and
  L.~Yang\inst{14} \and
  S.H.~Yong\inst{8}
}                     

\institute{
  Department of Physics, Sejong University, Seoul 05006, Republic of Korea \and
  Department of Physics, Yale University, New Haven, CT 06520, USA \and
  Physics Institute, University of S\~{a}o Paulo, 05508-090, S\~{a}o Paulo, Brazil \and
  Department of Physics and Astronomy, Seoul National University, Seoul 08826, Republic of Korea \and 
  Korea Institute of Science and Technology Information, Daejeon, 34141, Republic of Korea  \and
  Department of Physics, Bandung Institute of Technology, Bandung 40132, Indonesia \and
  Department of Physics and Astronomy, University of Sheffield, Sheffield S3 7RH, United Kingdom \and
  Center for Underground Physics, Institute for Basic Science (IBS), Daejeon 34047, Republic of Korea \and
  Department of Science Education, Ewha Womans University, Seoul 03760, Republic of Korea \and 
  Department of Physics, Sungkyunkwan University, Seoul 16419, Republic of Korea \and
  Department of Physics and Wisconsin IceCube Particle Astrophysics Center, University of Wisconsin-Madison, Madison, WI 53706, USA \and
  Department of Physics, Kyungpook National University, Daegu 41566, Republic of Korea \and
  Korea Research Institute of Standards and Science, Daejeon 34113, Republic of Korea \and
  Department of Physics, University of Illinois at Urbana-Champaign, Urbana, IL 61801, USA
}

\date{Received: date / Revised version: date}

\abstract{
  COSINE is a dark matter search experiment based on an array of low background NaI(Tl) crystals
  located at the Yangyang underground laboratory.
  The assembly of COSINE-100 was completed in the summer of 2016 and the detector is currently
  collecting physics quality data aimed
  at reproducing the DAMA/LIBRA experiment that reported an annual modulation signal.
  Stable operation has been achieved and will continue
  for at least two years.
  Here, we describe the design of COSINE-100, including the shielding
  arrangement, the configuration of the NaI(Tl) crystal detection elements, the veto systems, and
  the associated operational
  systems, and we show the current performance of the experiment.
} 

\maketitle

\section{Introduction} \label{intro}
Although dark matter appears to be ubiquitous, little is known about it.
Numerous astronomical observations, including the velocities of stars and galaxies,
anisotropies in the cosmic microwave background, and gravitational lensing measurements have indicated that
about 27\,\% of the Universe is comprised of dark matter~\cite{Huterer2010,Clowe:2006eq,Ade:2015xua}.
Theoretical physicists have proposed a Weakly Interacting Massive Particle (WIMP) as
a particle candidate for the dark matter~\cite{PhysRevLett.39.165,Jungman:1995df}.
They have suggested that rarely occurring interactions between WIMPs in the
Milky Way's dark matter halo and nuclei of normal matter may be measurable with a low-radioactivity detector
in a deep underground laboratory~\cite{Goodman:1984dc}.

One WIMP signature would be a modulation in the nuclear recoil event rate as an Earth-bound detector
sweeps through the galaxy's dark matter halo~\cite{PhysRevD.33.3495,Freese:2012xd}.
In a series of measurements that started in 1995, the DAMA/NaI and DAMA/LIBRA experiments (DAMA for short)
searched for evidence of an annual modulation signature in an array of low-background NaI(Tl)
crystals~\cite{Bernabei:2008yh}. Throughout this search,
the DAMA group has consistently reported a positive annual modulation signal with a phase consistent with expectations
for the Earth's motion relative to the galactic rest frame~\cite{bernabei00,Bernabei:2013xsa}.
Their most recent result, based on a 1.33\,ton$\cdot$year data with a 1\,count/day/kg/keV~\footnote{keV is electron equivalent energy.} background level crystal
array, is a 9.3\,$\sigma$ modulation in the single-site distribution of events in the 2 to 6\,keV range
with an amplitude of 0.0112$\pm$0.0012 count/day/kg/keV, a phase of 144$\pm$7\,days and a period of 0.998$\pm$0.002\,years.
 
The DAMA signal and, in particular, its interpretation as being due to WIMP-nucleon scattering, is
a subject of continuing debate~\cite{Freese:2012xd,bernabei00,Nygren:2011,Savage:2006qr,Kopp:2009et,Ralston:2010bd}.
This is primarily because WIMP-nucleon cross sections inferred from the DAMA modulation in the context of
the standard galactic WIMP halo model~\cite{Freese:2012xd} are in conflict with upper limits from other
experiments that are based on time-integrated measurements of the total rate of nuclear recoils, such as
LUX~\cite{PhysRevLett.118.021303}, PandaX~\cite{Tan:2016zwf,Cui:2017nnn}, XENON~\cite{aprile12,Aprile:2017iyp}, SuperCDMS~\cite{agnese14,PhysRevLett.116.071301,PhysRevLett.112.241302}
and KIMS~\cite{sckim12}. In addition, XMASS~\cite{Abe:2015eos} and XENON~\cite{PhysRevLett.118.101101} have reported that their annual modulation measurements based on leptophillic models are inconsistent with DAMA's results. However, no independent experimental confirmation of the DAMA signal with the
same target material and the same method has been performed to date.

The dark matter search region of interest corresponds to electron-equivalent nuclear recoil energies below 10\,keV.
In this energy region, the major internal background contributions in NaI(Tl) crystals are
\pbtwoten $\beta$ decays with Q$_\beta$=63.5\,keV, $\sim$3\,keV emissions from \kforty decays, and low energy emissions from cosmogenically
induced radioisotopes, including $\sim$3\,keV events from $^{109}$Cd, $\sim$4\,keV events from Te/I, and $^3$H beta decays~\cite{Bernabei:2008yh,adhikari16,Amare:2016rbf,Adhikari:2017gbj}.
The \pbtwoten contamination consists of a bulk contamination
that is primarily due to impurities in the raw materials used to produce the crystals and surface
contamination introduced by exposure to atmospheric $^{222}$Rn during crystal production and handling.
The bulk component of \pbtwoten(T$_{1/2}$=22.2\,yr) is difficult to remove from the raw material by commonly used purification methods.
The $\sim$3\,keV K-shell X-rays and Auger electrons from \kforty are mainly produced in the 10.55\% of the decays that proceed via electron capture to $^{40}$Ar with the emission of an accompanying 1460\,keV $\gamma$-ray.
The chemical similarity with sodium complicates the removal of potassium contamination from the NaI powder
used for the crystal growing.

One reason for the lack of verification of the DAMA result is that a new NaI(Tl) WIMP search would
require an independent development of low-background crystals. The crystal-growing
company that supplied the DAMA NaI(Tl) crystals no longer produces similar-grade crystals. 
Several groups including ANAIS~\cite{amare14A,amare14B}, DM-Ice~\cite{dmice,deSouza:2016fxg}, 
KamLAND-PICO~\cite{kamlandpico}, SABRE~\cite{sabre}, and KIMS~\cite{adhikari16,kykim15}, have worked to develop low-background NaI(Tl) crystals suitable for reproducing the DAMA experiment. 

Among these groups, KIMS and DM-Ice have joined to construct and operate a single experiment,
named COSINE, at the Yangyang underground laboratory (Y2L) in Korea.
KIMS and DM-Ice have assembled an eight element, 106\,kg array of low-background NaI(Tl) crystals that is currently
being used in the COSINE-100 experiment.
The construction and assembly of the COSINE-100 detector at Y2L took place in early 2016, and the
physics run started in late September of 2016. 

This paper is organized as follows:
Sec.~\ref{exparea} describes the experimental area and the detector room;
Sec.~\ref{shield} discusses the shielding arrangement;
Sec.~\ref{xtals} gives details about the internal radioactivity levels of the individual
                 crystals and describes how they are assembled into the detector array;
Sec.~\ref{lsvs} provides details about the liquid scintillator veto system, including results from prototype tests;
Sec.~\ref{psps} provides an overview of the cosmic-ray muon tagging system;
Sec.~\ref{daqdaq} gives a brief overview of the data acquisition system;
Sec.~\ref{slowslow} describes the environmental monitoring system;
Sec.~\ref{datadata} reports on the performance levels of the detector system and quality of the initial physics data;
Sec.~\ref{simu} provides descriptions of simulations;
Sec.~\ref{sensi} describes the expected sensitivity of COSINE-100 measurements; and
Sec.~\ref{conc} contains concluding remarks and comments.
\section{Experimental Hall} \label{exparea}
\begin{figure*}[!htb]
  \begin{center}
      \includegraphics[width=0.9\textwidth]{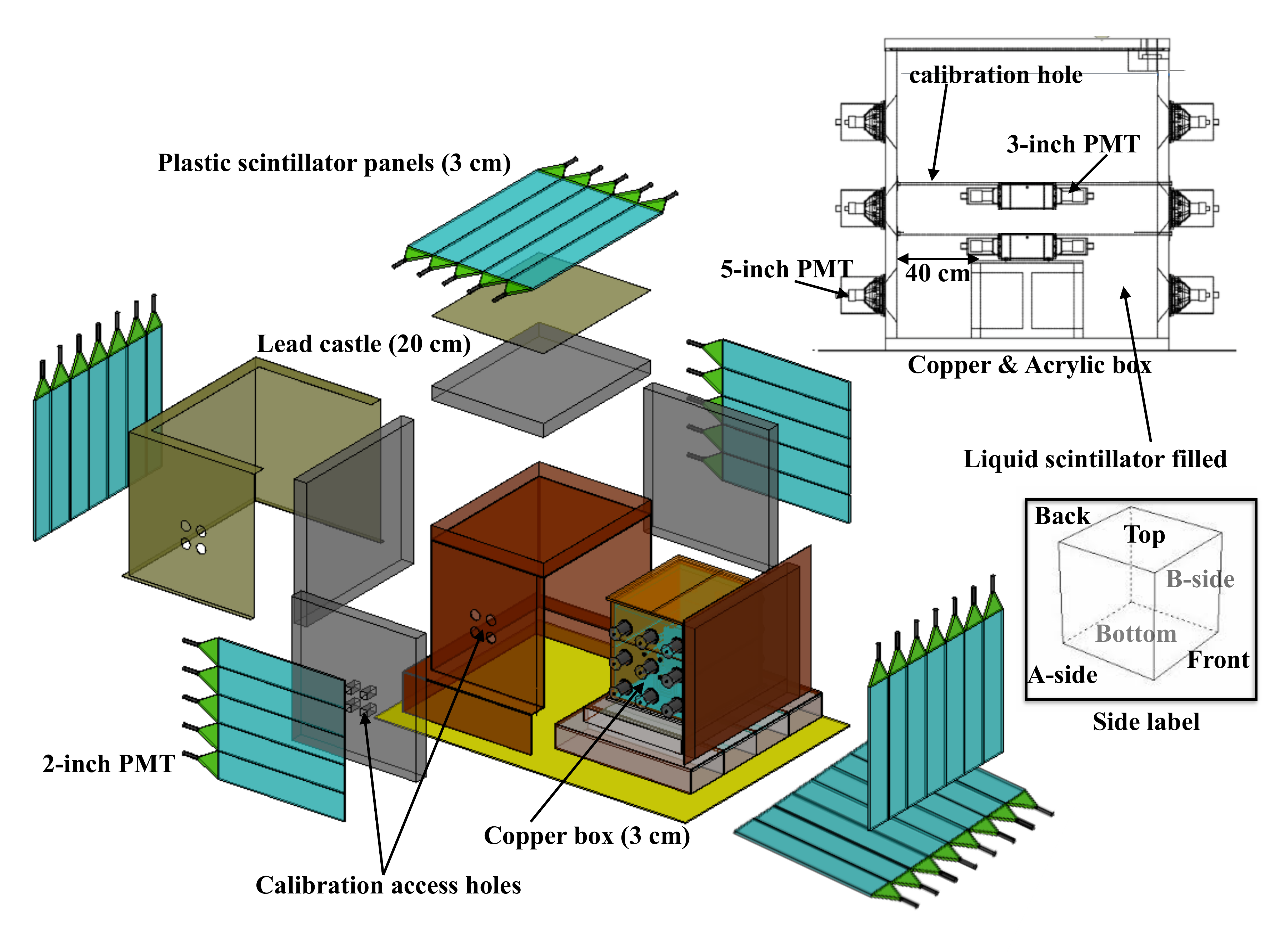}
  \end{center}
  \caption{COSINE shielding overview.
    From outside inward,
    muon panels (3\,cm thick, light blue),
    a lead brick castle (20\,cm thick on all sides, grey),
    a copper box (3\,cm thick),
    an acrylic box (1\,cm thick) and
    eight encapsulated crystal detectors immersed in the liquid scintillator ($>$40\,cm from crystal assembly to wall on all sides)
    are shown.
    Also indicated are the locations of the calibration holes, the size of the PMTs, and labeling scheme for the different sides.
    Here, the lead shields at the bottom and the front side are omitted for clarity.
  }
  \label{ref:shield}
\end{figure*}
\subsection{Experimental site at the Yangyang underground laboratory}
The COSINE experiment is located in a recently established experimental area in the Y2L A5 tunnel.
The Y2L facility is situated next to the underground generators of the Yangyang pumped-storage
hydroelectric power plant under Mount Jumbong, 150\,km east of Seoul in Korea
($\rm 38^{\circ}01'09.1"N$, $\rm128^{\circ}29'58.6"E$).
The laboratory consists of experimental areas located in the A5 and A6 tunnels
and are accessible by car via a 2\,km horizontal access tunnel. The experimental areas have a minimum granite overburden of
700\,m: the cosmic-ray flux in A5 is measured to be $\rm 3.80\pm0.01(stat.)\pm0.12(syst.)\times10^{-7}\,cm^{-2} s^{-1}$~\cite{Prihtiadi:2017inr}
while the flux in A6 is $\rm 2.7\times10^{-7}\,cm^{-2}s^{-1}$~\cite{muon_zhu}. 
Experimental rooms are built in the caverns located at the mid-sections of the tunnels. Automatically
regulated electrical power, conditioned by uninterruptible supplies, is provided to both experimental areas,
with voltages that are continuously monitored.
Fresh air from the surface is drawn into the tunnels through the driveways and exhausted via
a separate duct.  Throughout the year, the A5 tunnel temperature is maintained between 22\,$^\circ$C and
25\,$^\circ$C and the relative humidity near the laboratory rooms is measured to be in the 60--70\,\% range.

The Korea Invisible Mass Search (KIMS) experiment~\cite{sckim12}, which operated a CsI(Tl) array for dark matter searches in the A6 area for more than 15 years, has been discontinued and its shielding arrangement refurbished to host a variety of measurements as part of R\&D activities related to the development of low background detectors.

\subsection{Detector room}
COSINE-100 is located inside an environmentally controlled room with regulated humidity and temperature,
radon-reduced air, and gas supply systems that are monitored remotely online and in the surface-level
control room.
The detector room is 44\,m$^2$ in area and 4\,m high, and is maintained as an
access-controlled clean air environment.
In order to minimize contact with the air in the tunnel,
which contains 1.20$\pm$0.49\,pCi/L of \rntwotwotwo and other background components~\cite{Lee:2011jkps},
the room atmosphere is isolated from that of the tunnel.
The room air is continuously circulated through a HEPA filter and
the maximum number of dust particles larger than 0.5\,$\mu$m is kept below 1500 per cubic foot.
During detector installation periods, radon-reduced air, with a contamination that is a factor of 100
below the tunnel atmosphere level, is provided to the room. 
The air control system  maintains the room temperature at (23.5$\pm$0.3)\,$^\circ$C and the
relative humidity at (40$\pm$3)\,\%.
\section{Main shielding structure} \label{shield}
\begin{figure}[!htb]
  \begin{center}
      \includegraphics[width=0.48\textwidth]{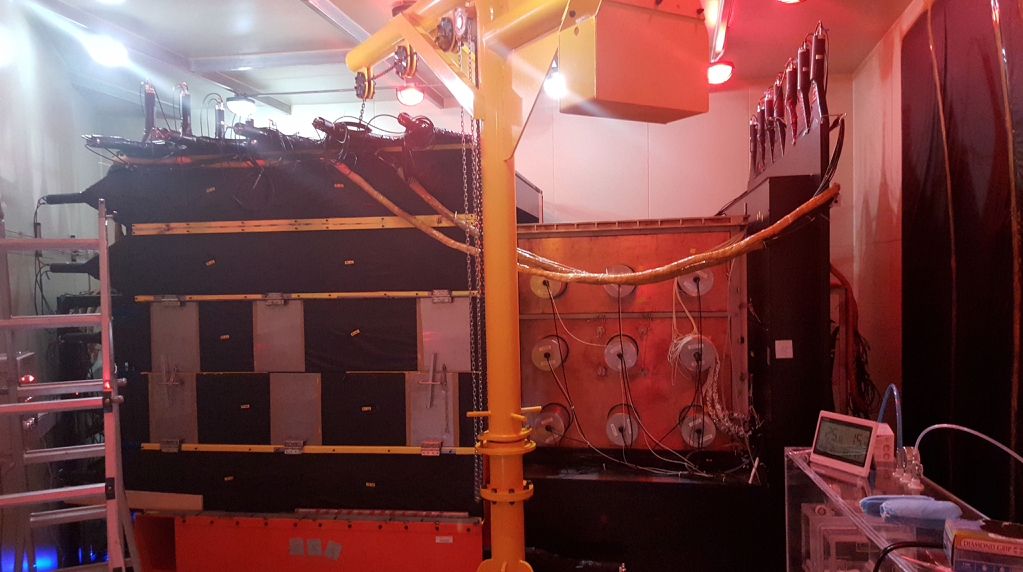}
  \end{center}
  \caption{Photo of the COSINE detector with the front side open.
    Plastic scintillators with black covers and an orange copper box with 5-inch PMT assemblies are shown. A hoist is used to move the 800\,kg top cover of the copper box.
  }
  \label{ref:room}
\end{figure}

The COSINE shielding structure inherited many features from KIMS~\cite{kims_crys1,Lee:2005qr}
and was designed to attenuate or tag the influence of external sources of radiation as efficiently
as possible. The detector is contained in a 4-layer nested arrangement of shielding components as shown
in Fig.\,\ref{ref:shield}. It provides 4\,$\pi$ coverage against external radiation from various sources. 
The shield is supported by a steel skeleton~\footnote{The steel has \utwothirtyeight and \thtwothirtytwo contamination
levels that were measured with the Y2L HPGe setup to be 40\,ppt and 100\,ppt, respectively.} that surrounds a
300\,cm\,(L)\,$\times$\,220\,cm\,(W)\,$\times$\,270\,cm\,(H) volume.  From outside inward, the four shielding layers are plastic
scintillator panels, a lead-brick castle, a copper box, and a scintillating liquid, as described below.
The eight NaI(Tl) crystal assemblies and their support table are immersed in the scintillating liquid.
The front side of the shield rests on a linear rail and can slide open at a speed of 40\,cm per minute.
A photograph of the detector with the front side open is shown in Fig.\,\ref{ref:room}. 

\subsection{Plastic scintillator panels}
An array of plastic scintillation counters provides
a simple and reliable method for tagging cosmic-ray muons that pass through
or near the detector.  The array of plastic scintillator panels that surrounds the COSINE detector
records and flags cosmic-ray muons and muon-induced events in the same data stream as the crystal data.
The main purpose of this system is to enable studies of correlations between cosmic-ray muons and crystal signals,
as high-energy muons are known to affect the response of NaI(Tl) crystals over a time interval
that extends beyond the 8\,$\mu$s window of the crystal readout system~\cite{Nygren:2011,Cherwinka:2015hva}.
Additionally, coincidence rates
between different panels in the system can be used to infer the cosmic-ray muon flux.

\subsection{Lead castle}

A 20\,cm-thick lead castle that surrounds the copper box attenuates $\gamma$-rays that originate from outside the shielding structure.
The inner half of this shield is made of low-contamination lead that
contains \utwothirtyeight and \thtwothirtytwo concentrations of 6.9\,ppt and 3.8\,ppt, respectively.
The $^{210}$Pb content was measured to be $30\pm1$\,Bq/kg at Y2L using alpha counting of $^{210}$Po decay
while the supplier's specification shows $59\pm6$\,Bq/kg using $^{210}$Bi beta counting.
The discrepancy is mainly due to different treatment of sample's surface emission.
The outer half of the shield is made of normal lead with 99.99\,\% purity.
The lead is in the form of 20\,cm\,$\times$\,10\,cm\,$\times$\,5\,cm rectangular bricks
stacked in such a way that there is no open channel between the outer and inner layers.
The total weight of the lead shield is 56\,tons.
The bricks were cleaned with ethanol prior to installation.

\subsection{Copper box}
The copper box serves as a shield for $\gamma$-rays as well as a support for the liquid scintillator.
ICP-MS measurements of the copper were 27\,ppt and 51\,ppt of \utwothirtyeight and $^{232}$Th, respectively.
The outer dimensions of the box are 152\,cm\,(L)\,$\times$\,142\,cm\,(W)\,$\times$\,142\,cm\,(H).
The wall thickness is 3\,cm and the total mass is 6.4\,tons. It is made of oxygen-free copper (OFC).
A 1\,cm-thick acrylic container for the liquid scintillator is nested inside of the copper box.

\subsection{Liquid scintillator}
A variety of backgrounds produced by radiogenic particles from components in and near the NaI(Tl) crystals, 
including the crystal PMT-originating and the NaI(Tl) internal backgrounds,
are efficiently rejected by an anticoincidence requirement with PMT signals from the liquid scintillator (LS)
and neighboring crystal signals.
This innermost active and passive shielding is provided by 2200\,L of Linear Alkyl-Benzene (LAB)-based LS 
contained in the acrylic box.
The inner walls of the acrylic container and the outer surfaces of the crystal assemblies
are wrapped with specular reflective films~\footnote{3M Vikuiti-ESR film} to increase the LS light collection efficiency.
The LS-produced photons are detected by eighteen 5-inch Hamamatsu PMTs\,(R877) that are attached to the two sides of the box.
The minimum distance between the crystal PMTs and the copper-box inner wall is approximately 40\,cm,
as indicated in Fig.\,\ref{ref:shield}. 

Through $\alpha$-particle measurements discussed in Sec.~\ref{lsvs}, we determined upper limits
for the intrinsic \utwothirtyeight
and \thtwothirtytwo contaminations in the LAB-LS of 7\,ppt and 4\,ppt, respectively.

The top 9\,cm of the acrylic box holding the LS was left unfilled as a safety margin in the event of a
temperature increase that might cause an expansion of the LS volume.
To maintain a high LS light output by preventing contact with oxygen and water, gas boil-off from a liquid nitrogen dewar
is supplied to this space at a rate of 3 liters per minute and the humidity at the top of the liquid is maintained at $<$2.0\,\%.
The scintillating liquid volume and its relatively high heat capacity helps keep the temperature of the liquid and crystals stable at (24.20$\pm$0.05)\,$^\circ$C.
\section{NaI(Tl) Crystal Detectors} \label{xtals}
\subsection{Crystal assembly}
The COSINE-100 experiment uses low-background NaI(Tl) crystals (labeled Crystal-1 to Crystal-8, or C1$-$C8)
that were developed in cooperation with Alpha Spectra Inc.\,(AS).
The eight NaI(Tl) crystals were grown out of batches of powders with successive improvements,
with AS-B and AS-C: Alpha Spectra purified powders, AS-WSII: Alpha Spectra WIMPScint-II grade powder and AS-WSIII: Alpha Spectra WIMPScint-III grade powder.
The final crystals are cylindrically shaped and hermetically encased in OFC tubes (1.5\,mm thick) with quartz windows attached at each end.
Each crystal's lateral surfaces were wrapped in roughly 10 layers of 250\,$\mu$m-thick PTFE reflective sheets and then inserted into the OFC tubes in a nitrogen gas environment and sealed to make them gas tight.
A 12.0\,mm-thick quartz window is light-coupled to each end of the crystal via 1.5\,mm thick optical pads~\footnote{Eljen Technology EJ-560}.
These, in turn, are light-coupled to 3-inch Hamamatsu R12669SEL PMTs
via a small amount of high viscosity optical gel~\footnote{Eljen Technology EJ-550}. The average quantum efficiency of the PMTs is 35\,\%.
The copper cylinders that encapsulate the crystals have 
16\,mm-diameter calibration windows
with either a reduced copper thickness of 0.5mm or a
0.13mm-thin Mylar cover to facilitate low energy source calibrations.

Four different powder grades were used to grow the eight crystals (see Table~\ref{activity}).
Although AS does not release the specifics of each powder for proprietary reasons, results from
our detailed studies of these powders and crystals are described in Refs.~\cite{adhikari16,kykim15}.
\begin{table*}[ht]
  \caption{Measured radioactivity levels in the COSINE-100 crystals.
    The light yield is measured at 59.6\,keV with a \amtwofortyone source and checked
    for consistency with the 46.5\,keV internal \pbtwoten $\gamma$-ray peak.
    The units for light yield are photoelectrons per keV (PEs/keV).
    Chain equilibrium is assumed for the \utwothirtyeight and \thtwothirtytwo values.
  }
\label{activity}
\begin{tabular}{lcccccccc}
  \hline                                                                                
  Crystal    & Mass & Size (inches     & Powder    & $\alpha$ Rate & \kforty  & \utwothirtyeight & \thtwothirtytwo  &  Light Yield        \\
             & (kg) & diameter$\times$length)     &           & (mBq/kg) & (ppb)  & (ppt)  & (ppt)  &(PEs/keV)           \\
  \hline                 
  Crystal-1  & 8.3  & $5.0\times7.0$     &  AS-B     & $3.20\pm0.08$  & $34.7\pm4.7$  & \textless0.02  & $1.3\pm0.4$   &  $14.9\pm1.5$    \\
  Crystal-2  & 9.2  & $4.2\times11.0$     &  AS-C     & $2.06\pm0.06$ & $60.6\pm4.7$   & \textless0.12  & \textless0.6  &  $14.6\pm1.5$   \\
  Crystal-3  & 9.2  & $4.2\times11.0$     &  AS-WSII  & $0.76\pm0.02$ & $34.3\pm3.1$    & \textless0.04  & $0.4\pm0.2$   &  $15.5\pm1.6$     \\
  Crystal-4  & 18.0 & $5.0\times15.3$     &  AS-WSII  & $0.74\pm0.02 $ &$33.3\pm3.5$    &                & \textless0.3  &  $14.9\pm1.5 $       \\
  Crystal-5  & 18.3 & $5.0\times15.5$     &  AS-C     & $2.06\pm0.05 $ & $82.3\pm5.5$  &                & $2.4\pm0.3$   &  $7.3\pm0.7 $     \\
  Crystal-6  & 12.5 & $4.8\times11.8$     &  AS-WSIII & $1.52\pm0.04 $ & $16.8\pm2.5$   & \textless0.02  & $0.6\pm0.2$   &  $14.6\pm1.5 $      \\
  Crystal-7  & 12.5 & $4.8\times11.8$     &  AS-WSIII & $1.54\pm0.04 $ & $18.7\pm2.8$   &                & \textless0.6  &  $14.0\pm1.4 $    \\
  Crystal-8  & 18.3 & $5.0\times15.5$     &  AS-C     & $2.05\pm0.05  $  & $54.3\pm3.8$ &                & \textless1.4  &  $3.5\pm0.3  $     \\
  \hline                 
  DAMA       &      &          &           & $<0.5$ &  $<20$ & 0.7$-$10  & 0.5$-$7.5 &  5.5$-$7.5    \\
  \hline  
\end{tabular}
\end{table*}

\subsection{Calibration system and crystal array geometry}
The performance levels of the crystals and the liquid scintillator are monitored with
a variety of calibration sources;  four different $\gamma$-ray sources including $^{241}$Am,
$^{57}$Co, $^{137}$Cs and $^{60}$Co are used for the energy calibration.
Four stainless steel tubes with a 9.5\,mm outer-diameter and 2.5\,mm thickness penetrate the copper and acrylic boxes and pass through the liquid scintillator.
Access to the calibration tube requires the dismounting of two of the plastic scintillators and
the removal of eight of the lead castle bricks.  During calibration campaigns, needle-type source tips
encapsulated in a stainless steel rod are inserted into the tube.  In this way, calibrations can be done
without opening the front shield.

\begin{figure*}[!htb]
  \begin{center}
    \begin{tabular}{cc}
      \includegraphics[width=0.5\textwidth]{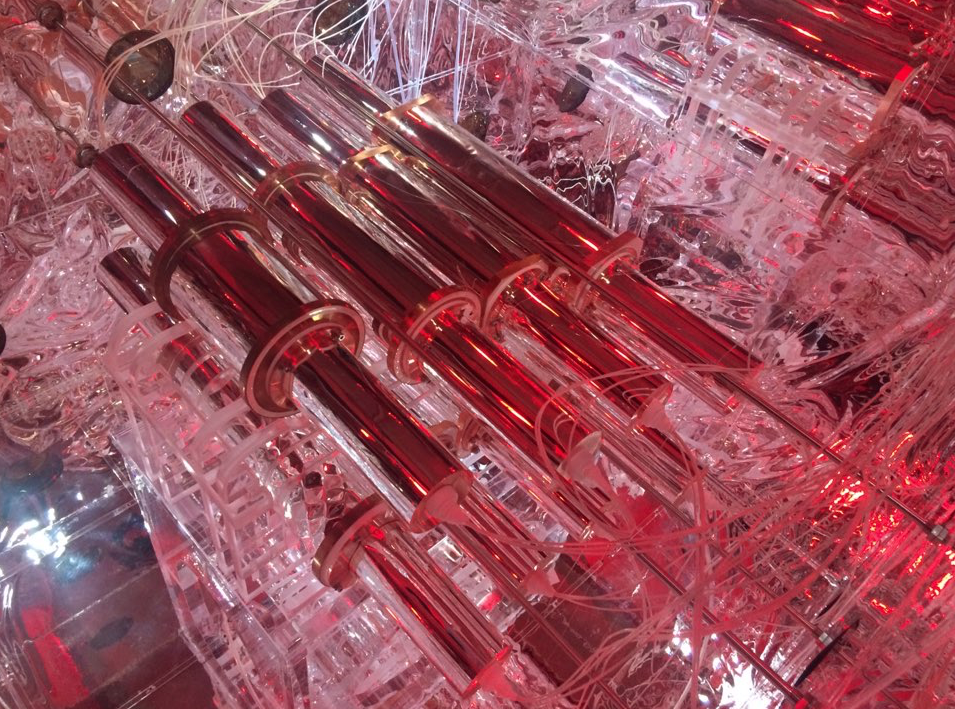} &
      \includegraphics[width=0.35\textwidth]{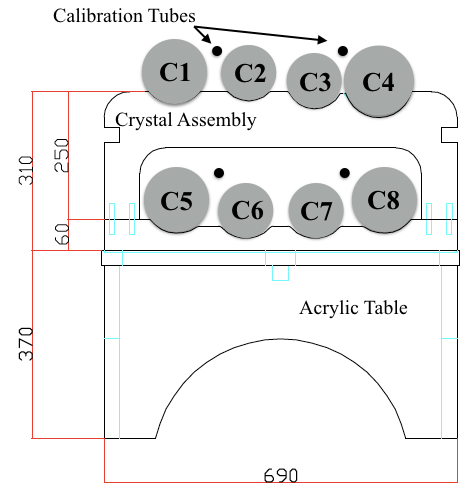} \\
      (a) & (b)\\
    \end{tabular}
  \end{center}
  \caption{The $4\times2$ NaI(Tl) crystal array.
    Each crystal has two, 3-inch PMTs and is encapsulated in
    a thin Polytetrafluoroethylene (PTFE) reflector and copper casing.
    The crystal arrangement is shown in (a) and a diagram of the crystal arrangement is shown in (b).
    An acrylic table supports the crystal assemblies.
    The eight crystals are labeled C1 through C8 and the locations of the four calibration holes are indicated.
    The table dimensions are indicated in mm and viewed from the A-side of the detector.
  }
  \label{ref:crystals}
\end{figure*}

The eight NaI(Tl) crystals are arranged in a 4$\times$2 array that is supported by
a two-layer acrylic table located in the central region of the liquid scintillator.
The crystals are oriented such that their calibration windows
directly face the nearest calibration tube.
Two external source calibrations were performed on September 22, 2016 and on December 27, 2016 so far.
The crystal arrangement and numbering scheme are shown
in Fig.~\ref{ref:crystals}.

\subsection{Background activity of the crystals}
Before the installation of the crystals in the COSINE shield,
background event rates from internal radioactive contaminants in the NaI(Tl) crystals were measured
in the KIMS-CsI setup at A6~\cite{adhikari16,kykim15}.  After the insertion of the crystals into the shield and prior to filling the liquid scintillator container, their background levels were
remeasured to verify that they were free of any additional contamination.
Overall, the eight crystals have high light yields and
acceptable \utwothirtyeight and \thtwothirtytwo contaminations (see Table~\ref{activity}).
The light yield of Crystal-5 and Crystal-8, however, decreased over the first few weeks
and their optical coupling is suspected to have degraded.

The $^{210}$Pb levels were improved by a factor of two between the AS-C and AS-WSII powders,
and \kforty levels also improved by a factor of two between the WS-II and WS-III powders.
The powder grade closely correlates with the contamination level of the grown crystals,
as demonstrated by the \kforty levels, which vary strongly according to the powder type;
within each specific batch of powder, the \kforty radioactivity levels are consistently reproduced.
The \potwoten(in equilibrium with $^{210}$Pb) levels, as determined from the $\alpha$ particle rates, are higher than those
achieved by DAMA~\cite{Bernabei:2008yh}; the origin of this elevated rate is not yet fully understood. 
Some of the \pbtwoten may originate from lead or radon that is introduced into the raw materials
prior to crystallization.
Chain equilibrium is assumed for the interpretation of \utwothirtyeight and \thtwothirtytwo
related radioactivity measurements, with the exception of $^{210}$Pb.

\section{Liquid Scintillator Veto system} \label{lsvs}
The LAB-LS contains a few percent of 2,5-Diphenyloxazole (PPO) as the
fluor and for wavelength shifting, with a trace amount of
p-bis-(o-methylstyryl)-benzene (bis-MSB) to serve as a second wavelength shifter~\cite{JS_Park_1}.
To produce the final 2200\,L  LAB-LS volume, a 110\,L concentrated solution of PPO and bis-MSB was first prepared.  This master solution was then mixed with LAB
in a 1:20 ratio to make the final LAB-based LS. 
The LS was produced in a surface-level laboratory and moved to the A5 tunnel at Y2L immediately
after production. The LS was passed through two layers of 0.1\,$\mu$m pore-size filters while being transferred
into the acrylic container.

Since PPO may have a relatively high level of contamination by radioisotopes,
a water extraction method~\cite{waterextract} with 17\,M$\Omega$ deionized water was used
to extract impurities from the master solution.  To remove the residual water after this treatment,
a high flux ($\sim$30 L/min) of N$_{2}$ gas was bubbled through the liquid 
until the exhaust-gas humidity level reached $\sim$20\,ppm $\rm H_2O$.

\begin{figure*}[!htb]
  \begin{center}
    \begin{tabular}{cc}
      \includegraphics[width=0.48\textwidth]{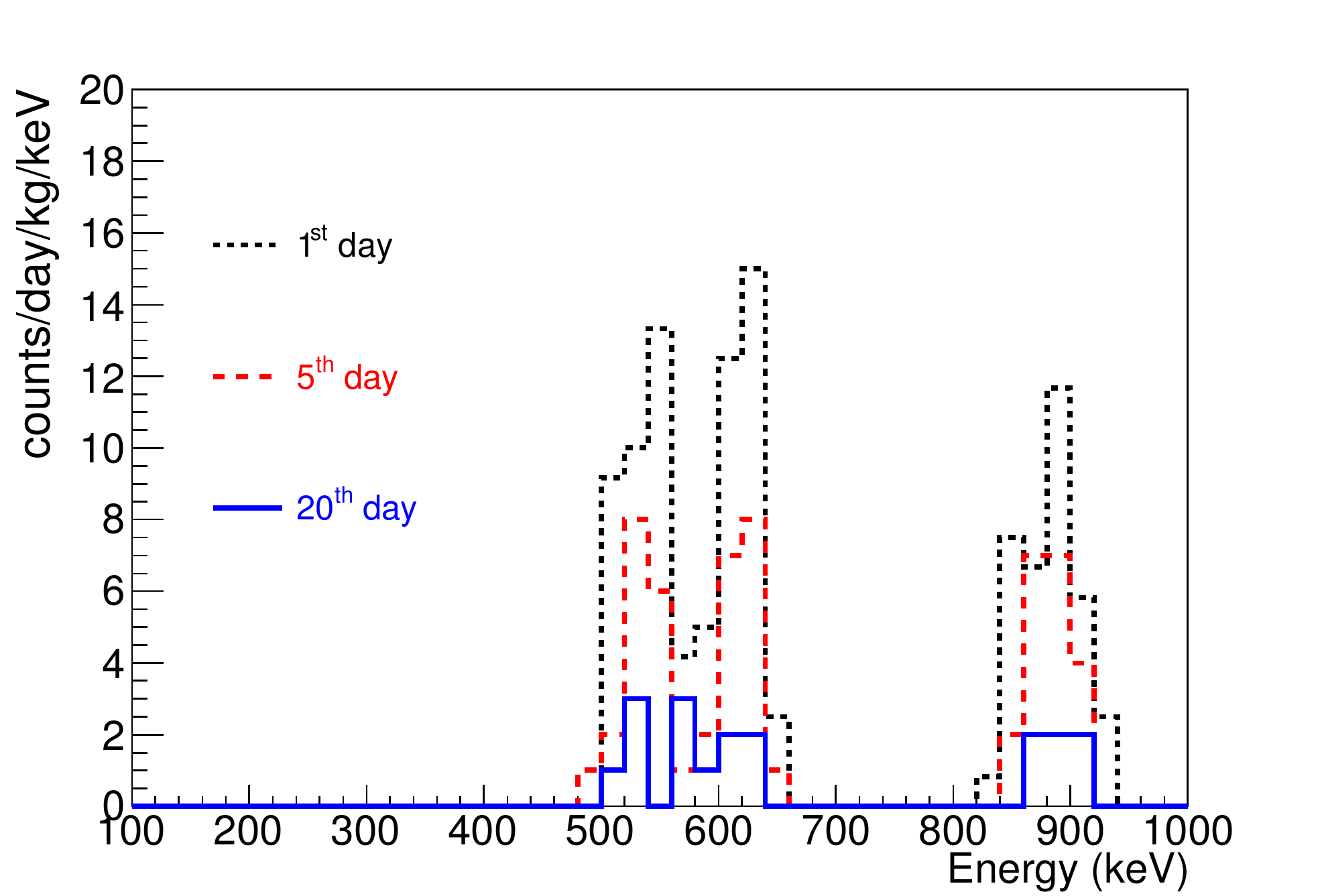} &
      \includegraphics[width=0.48\textwidth]{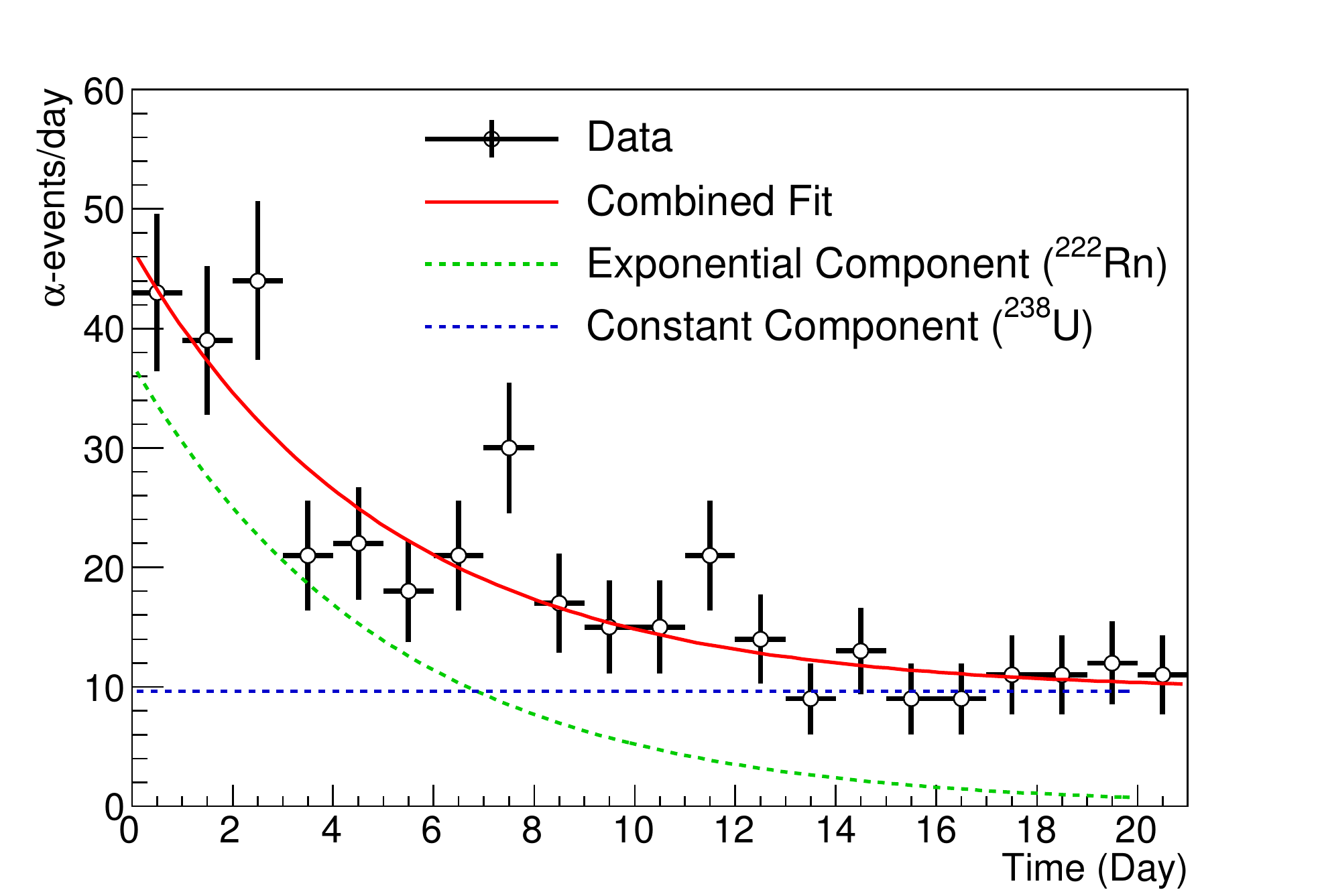}\\
      (a) & (b) \\
    \end{tabular}
  \end{center}
  \caption[]{
    (a) Energy spectra of $\alpha$-particles originating from the 70\,mL PTFE cell at different times.
    The histogram labeled ``$\mathrm{1^{st}}$ day'' is when the 70\,mL cell was first assembled while that
    labeled $\mathrm{20^{th}}$ day is 19 days later.
    The $\alpha$-induced events are selected based on pulse shape discrimination criteria.
    (b) The number of measured $\alpha$-events as a function of time.
    A fit with an exponential plus a flat component was performed with the assumption that decreasing components originate
    from external contamination of $^{222}$Rn and from its daughter particle decays, while the flat component is produced by long lived isotopes in the $^{238}$U chain that are intrinsic to the LS.
  }
  \label{fig:jsp_radon}
\end{figure*}

Internal $\alpha$-contamination measurements of the LAB-based LS were
carried out with a 70\,mL PTFE cell detector. Figure~\ref{fig:jsp_radon} shows the
results of these background measurements. Signals induced by $\alpha$ particles are separated from $\gamma$-induced events
by means of pulse shape discrimination.
The observed $\alpha$-energy spectrum is understood as \rntwotwotwo contamination
that might have occurred 
during the assembly of the small-cell detector. As can be seen in Fig.~\ref{fig:jsp_radon},
the $\alpha$-peaks decreased at a rate that is consistent with the 3.8\,day half life of $^{222}$Rn.
A simulation of \pbtwoten shows that
the remaining \pbtwoten in LS does not contribute a significant observable background in the crystals.
Because we only observe $\alpha$-events corresponding to $^{222}$Rn, the level of the intrinsic \utwothirtyeight chain
activity in the LAB-LS was determined to be 7\,ppt from a fit to the time dependence
with a flat component in addition to an exponential with a 3.8\,day half life.
An upper limit for an intrinsic \thtwothirtytwo chain activity was measured to be 4\,ppt using a time difference analysis of consecutive $\alpha$-decays~\cite{kykim15}.
Measurements with a prototype LS detector and simulation studies show that these upper limits are sufficiently small
to ensure negligible background contributions to the NaI(Tl) crystal energy spectra~\cite{Park:2017jvs}.

\begin{figure}[!htb]
\begin{center}
    \begin{tabular}{c}
      \includegraphics[width=0.5\textwidth]{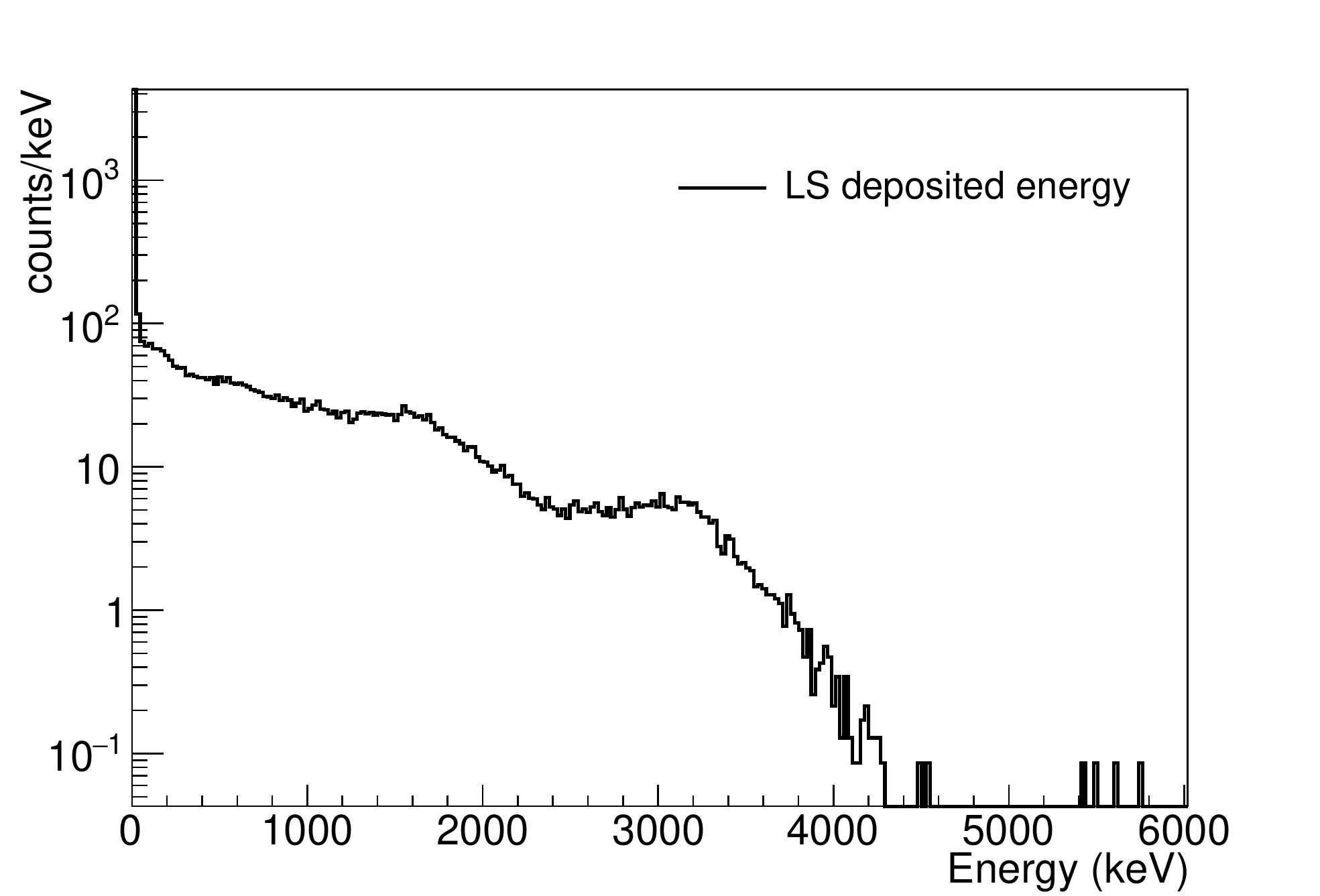}
    \end{tabular}
    \caption{The LS energy spectrum for initial 16.7 days of data.
      The signals from all 18 readout PMTs are summed.
      Note that the LS PMTs are not included in the event trigger. Only passive data
      that occur during a time window that includes the crystal trigger time are recorded.}
  \label{lsspectrum}
  \end{center}
\end{figure}

Energy calibrations of the LS veto system are performed with various $\gamma$-ray
sources inserted in the calibration tubes.
The position dependence of the light yields was inferred 
from measurements made with disk sources placed at different positions on the outside of the LS container during the commissioning period. After applying these LS characteristics to data, the LS deposited energy spectrum
shown in Fig.~\ref{lsspectrum} is obtained.

\begin{figure}[!htb]
\begin{center}
    \begin{tabular}{c}
      \includegraphics[width=0.5\textwidth]{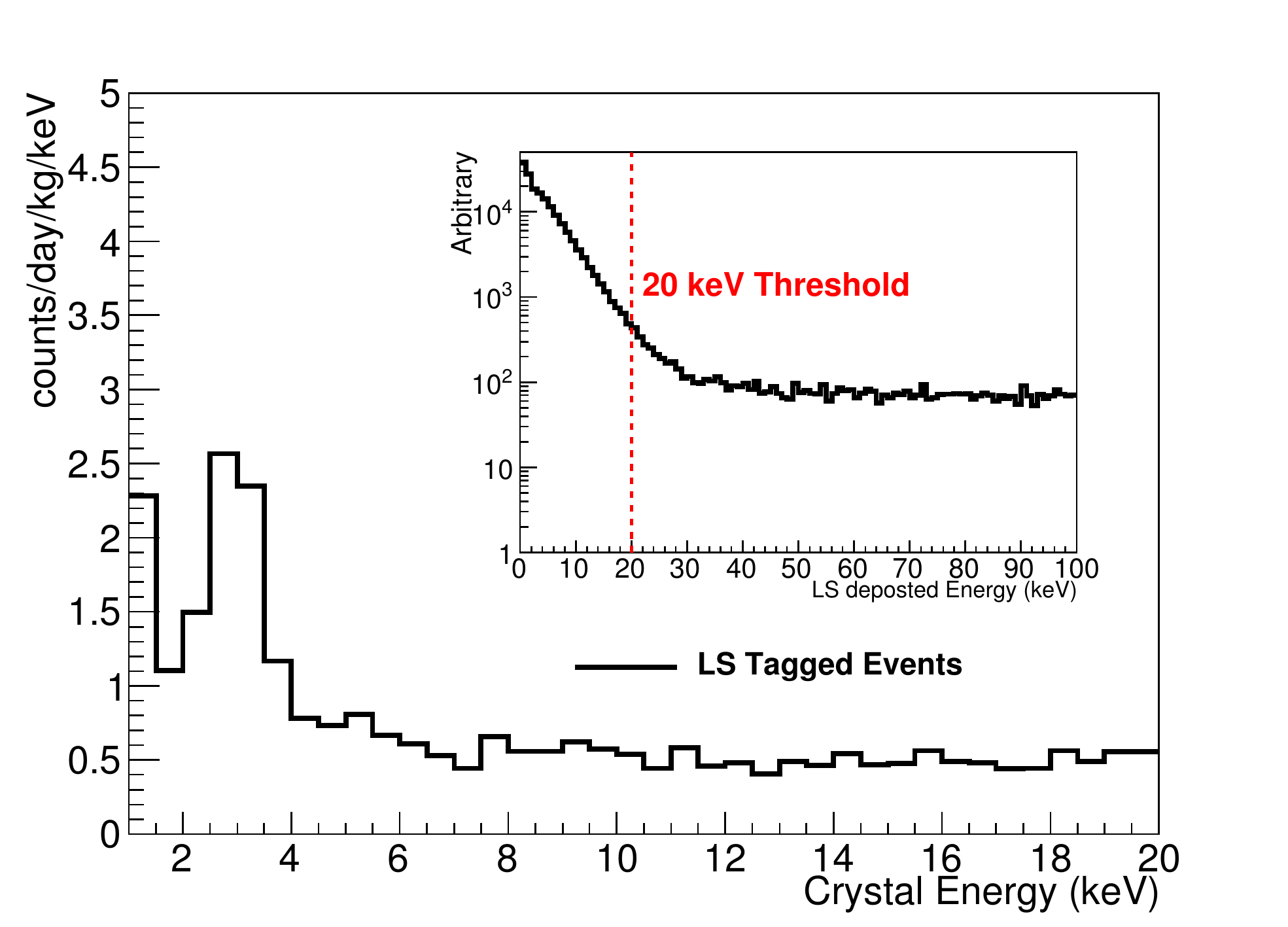}
    \end{tabular}
    \caption{
      The spectrum of Crystal-1 signals that occur in coincidence with LS signals
      with E$>$20\,keV is shown. The 3\,keV \kforty events are tagged in the
      LS system with a 72\% efficiency.
    }
  \label{ls}
  \end{center}
\end{figure}

Coincident events between the LS and a single crystal show a time correlation for LS signals with energies
above 20\,keV.  With a 20\,keV LS energy threshold requirement 
and a 200\,ns coincidence window between LS and crystal signals, there are
only 0.3\,\% accidental coincidences in the sample of the selected events. 
With these conditions, the energy spectrum of events in Crystal-1 that occur in coincidence with LS signals is shown in Fig.~\ref{ls}.
By comparing the crystal energy spectrum with and without the coincidence condition in the 2--4\,keV region,
we can determine a 72\,\% tagging efficiency for 3\,keV emissions from $^{40}$K decay.
Similarly, Crystal-2 has a tagging efficiency of 74\,\%.
Both agree well with simulated efficiencies and the measured $^{40}$K activities from the previous measurements in the KIMS array.
For other crystals, this data-driven efficiency estimation is difficult due to
additional 3\,keV events from cosmogenically produced $^{109}$Cd, low light yields or low internal $^{40}$K levels.
For these crystals, tagging efficiencies from simulations are between 65 and 75\,\%, depending on the crystal's location in the array.

\section{Muon detector system} \label{psps}
It is known that the cosmic-ray muon intensity is modulated by seasonal temperature variations of the
atmosphere~\cite{Cherwinka:2015hva,Bellini:2012te,PhysRevLett.105.121102}. Thus, signals in COSINE-100
that are induced by cosmic-ray muons could mimic dark matter annual modulation signals.
To tag cosmic-ray muons and study their correlations with crystal signals, we have surrounded
the COSINE-100 detector with 37 muon detectors made from panels of 3\,cm thick plastic scintillator.
The scintillation panels are read out by 2-inch Hamamatsu H7195 PMTs via acrylic light guides.

\subsection{Muon detector panels}
The muon detector system is made of scintillating plastic from Eljen Technology, EJ-200. 
This material provides favorable properties, such as high light output (10,000 photons/MeV), PMT friendly wavelengths (efficiency peaks at 425\,nm) and a long optical attenuation length (380\,cm).

\begin{figure}[!htb]
  \begin{center}
    \includegraphics[width=0.5\textwidth]{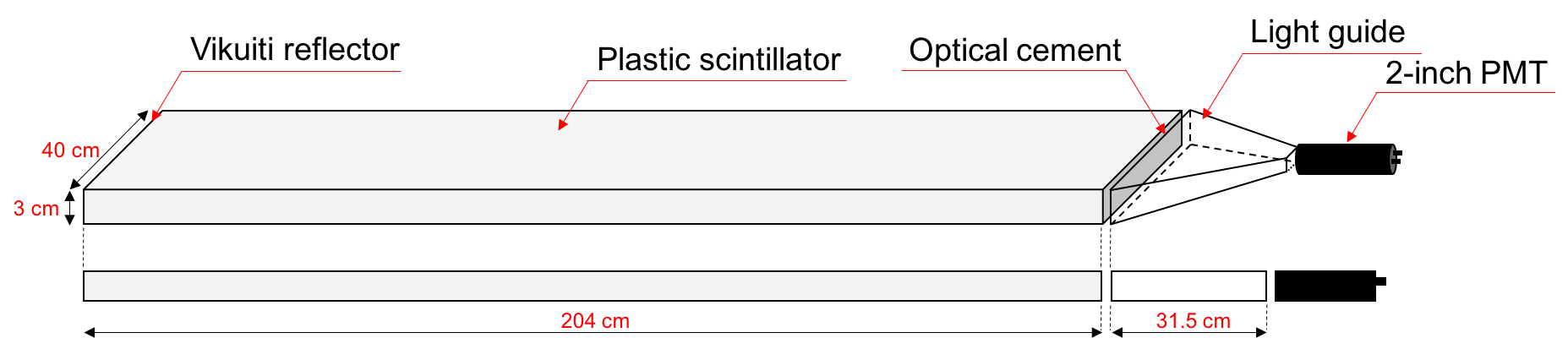}
  \end{center}
  \caption{A schematic view of a muon detector.}
  \label{muonpanelcartoon}
\end{figure}

\begin{figure}[!htb]
  \begin{center}
    \includegraphics[width=0.48\textwidth]{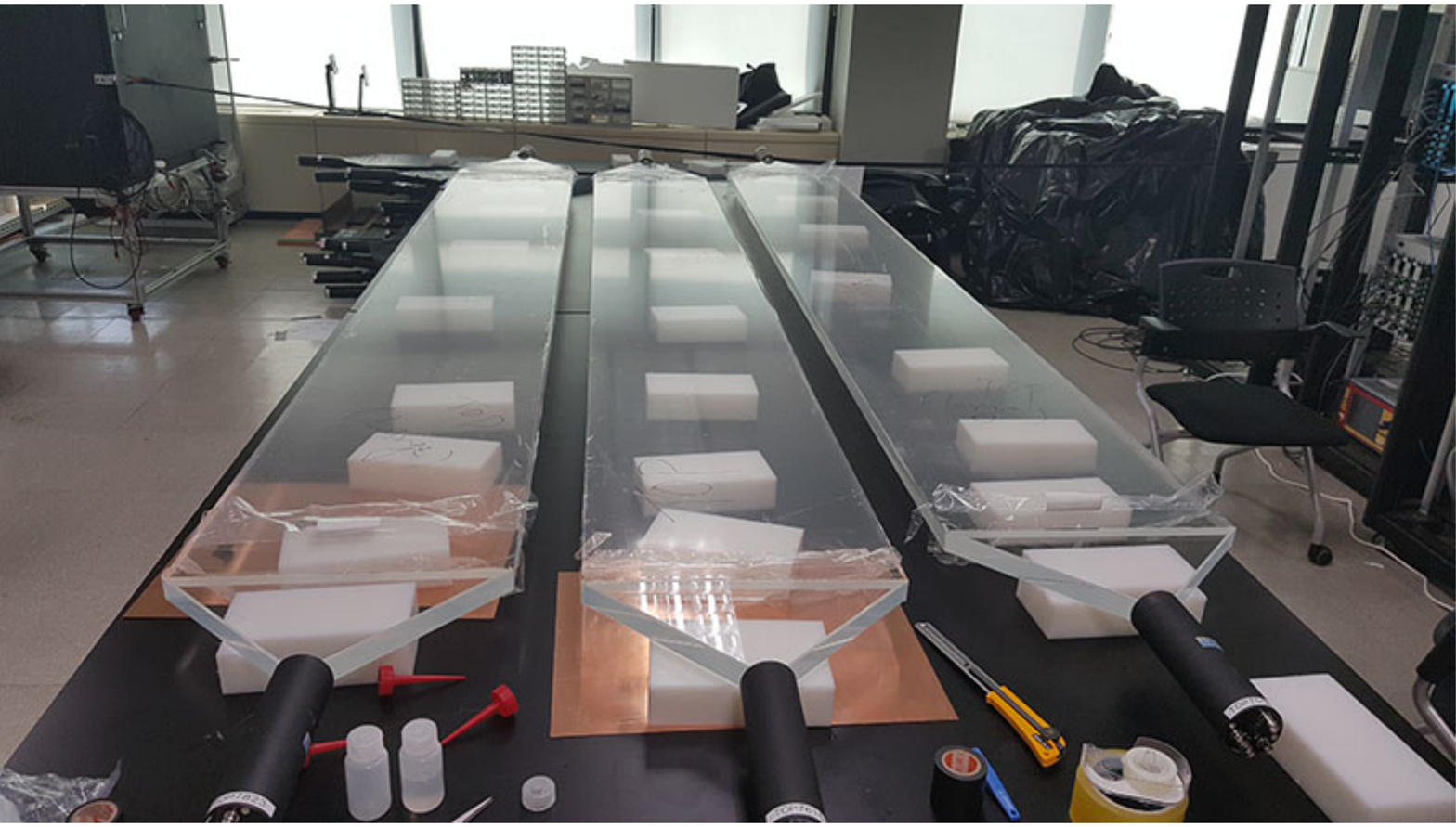}
  \end{center}
  \caption{A photo of the muon detector production.}
  \label{muonpanelproduction}
\end{figure}

Prior to attaching the PMTs, each scintillator was polished and
wrapped with white reflective sheets, a layer of aluminum foil, and
a black sheet, an arrangement that was found to provide efficient light collection,
create a barrier against external light leaking into the scintillator,
and protect against physical damage to the plastic.
Epoxy~\footnote{Saint Gobain, BC-600} is used to make the optical connection between the plastic scintillator and the light guide, and
between the light guide and the PMT photocathode.
For the 2\,m long panels that are coupled to only one PMT,
a specular reflecting sheet is attached to the edge of the scintillator that is opposite of the PMT.
This improves the light output by about 8\%.
The 3\,m long panels are coupled to a PMT at each end and form the top surface of the main shield.
The arrangement of the COSINE-100 muon detection panels in the shielding structure
is shown in Fig.~\ref{ref:shield}.
A schematic diagram of a single panel is shown in Fig.~\ref{muonpanelcartoon};
a production photo is shown in Fig.~\ref{muonpanelproduction}.

\subsection{Performance of the plastic scintillator}
We performed various tests on the panels in a surface-level laboratory where the cosmic-ray muon flux is high.
For these tests, a small-size muon panel placed above the muon detector under study provided a trigger signal.
This arrangement was used to determine the light yield and its dependence on the distance to the PMT.
To select muon candidates, we required a coincidence between the trigger panel and the panel being tested.
With a moderate threshold requirement, cosmic-ray muon candidates were selected and their charge spectrum 
was fitted to a Landau function.
The most probable value from the fit results was subsequently used to calibrate
the charge spectrum of a single panel as well as to align the panel-to-panel spectra.

To understand the performance and efficiency of each muon detector panel,
we stacked them four at a time and triggered on coincidences between pairs of counters.
By comparing the responses of one pair in events triggered by the other pair, we inferred
the panel efficiencies for cosmic-ray muons to be 99.5\%; the 0.5\% loss could be attributed
to horizontal muons that strike the edge of one panel and minor misalignments of the panels.

\begin{figure}[!htb]
  \begin{center}
    \includegraphics[width=0.48\textwidth]{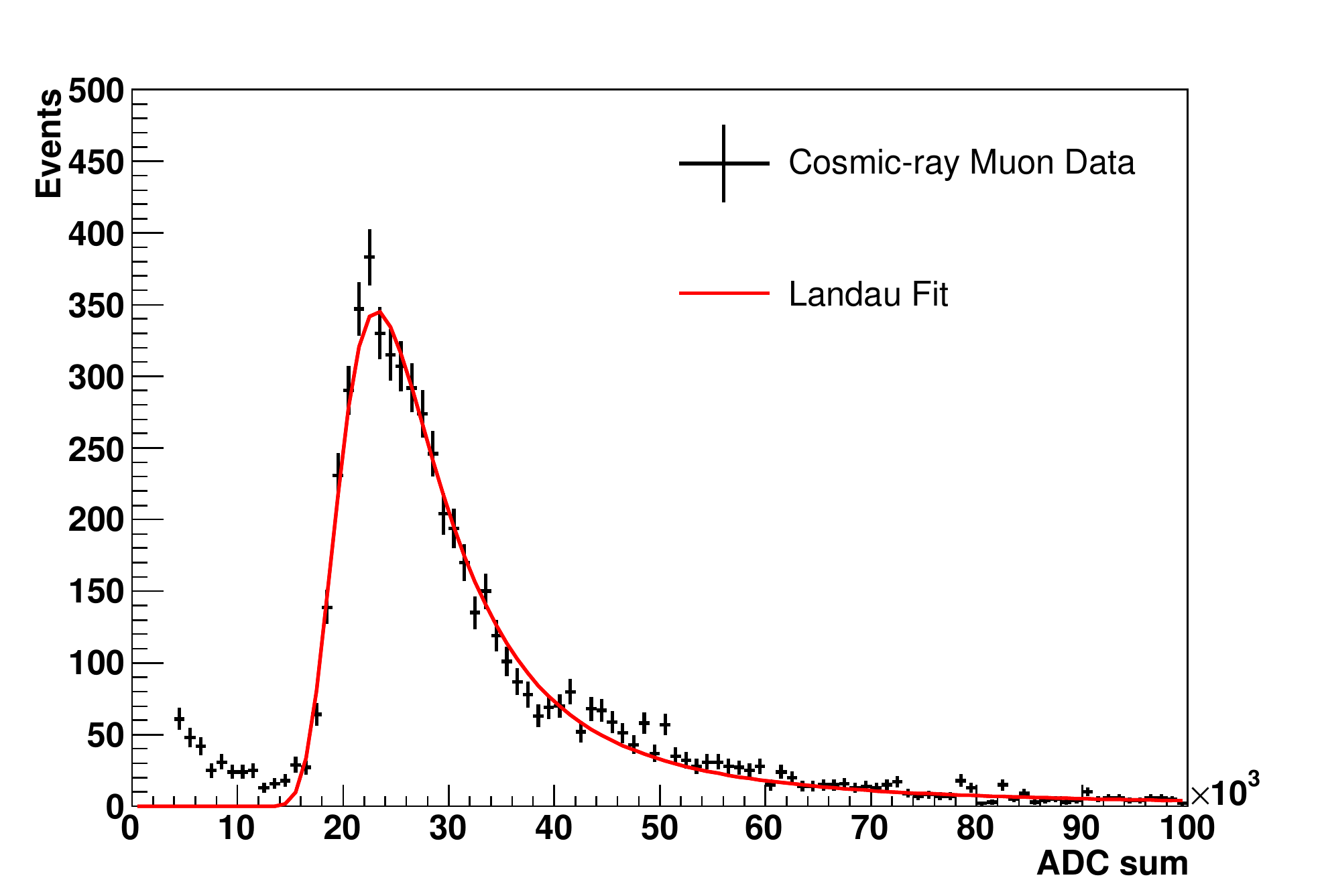}\\ 
    \caption{The summed charge spectrum of muon-like events in one of the top-side panels
      for 48.3 days of initial data in the COSINE-100 detector.
      The solid curve is the result of a fit to the spectrum with a Landau function.
      Events with ADC sum below 14,000 counts are primarily due to $\gamma$-ray-induced events.
    }
    \label{muonspectrum}
  \end{center}
\end{figure}

The muon detection panels were subsequently installed underground in the COSINE-100 shielding
structure at Y2L.
Figure~\ref{muonspectrum} shows the total charge distribution from one of the top panels where
muon-like events are well separated from low-energy $\gamma$-ray-induced backgrounds.
The cosmic-ray muon flux at the COSINE-100 detector
was measured to be 328$\pm$1(stat.)$\pm$10(syst.) muons/m$^2$/day~\cite{Prihtiadi:2017inr}.
The current NaI(Tl) crystal event selection requires a coincidence within a 30\,ms time window
between a crystal triggered event and a muon panel event to select the muon induced events.
A further refinement of this condition and studies of those selected events for their time dependence
are actively on-going.

\section{Electronics and data acquisition system} \label{daqdaq}
\subsection{Electronics and data flow}
\begin{figure*}
\begin{center}
\includegraphics[width=0.8\textwidth]{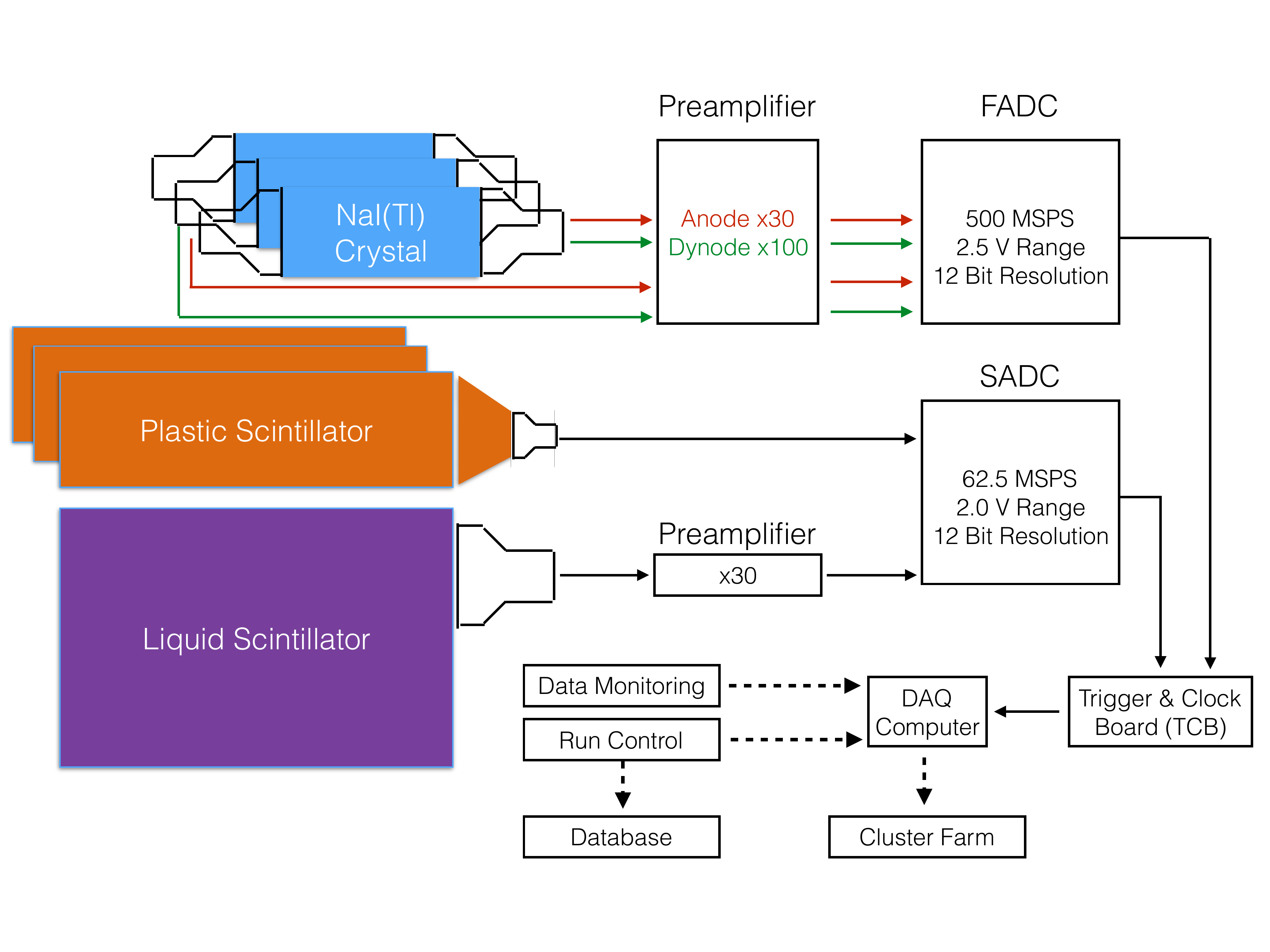}
\end{center}
\caption[]{Data flow block diagram. The crystal signals are recorded with FADCs while the plastic and liquid
scintillator signals are recorded with SADCs. Global triggers are formed at the TCB board.} 
\label{fig:jsp_data_flow}
\end{figure*}

The data acquisition system is located inside the detector room
in order to minimize the signal attenuation through cables and to reduce environmental effects.
The system consists of DAQ modules, HV supply modules, and computers.
All electronics and their AC power are controlled by a dedicated computer
in the control room adjacent to the detector room.
Figure~\ref{fig:jsp_data_flow} shows an overall data-flow diagram for the
COSINE-100 experiment.

There are 16 3-inch crystal-readout PMTs, 18 5-inch PMTs for the
LS system and 42 2-inch PMTs for the muon system.
Since each crystal PMT has two readout channels, there are a total of 92 signal channels that
are read out and recorded by the DAQ system.  In addition, there are a total of 76 separate high
voltages that have to be supplied and monitored.

\begin{figure*}
\begin{center}
  \includegraphics[width=0.95\textwidth]{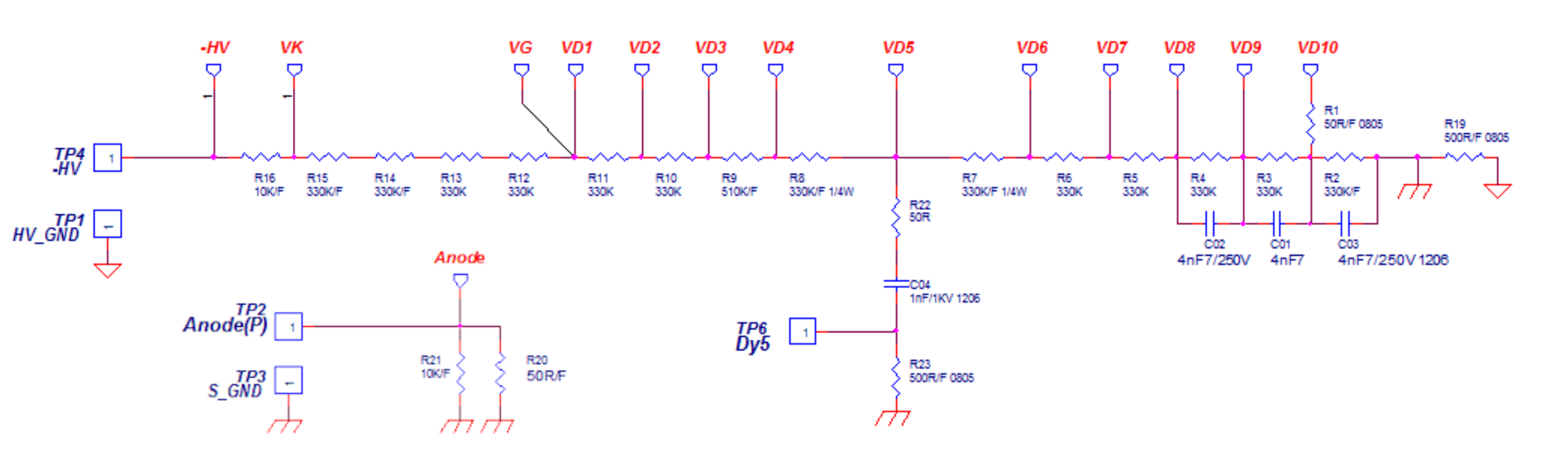}
\end{center}
\caption[]{Schematic of crystal PMT voltage divider.} 
\label{fig:vdivider}
\end{figure*}

Signals from each crystal PMT are amplified by custom-made preamplifiers:
the high-gain anode signal by a factor of 30 and
the low-gain signal from the 5$^{\rm th}$ stage dynode by a factor of 100.
Figure~\ref{fig:vdivider} shows the PMT voltage divider diagram where the anode and 5$^{\rm th}$ dynode readout circuits are
indicated.
The amplified signals
are digitized by 500\,MSPS\,(megasamples per second), 12 bit flash analog-to-digital converters\,(FADCs).
Both the low-gain dynode
and high-gain anode waveforms are recorded whenever an anode signal produces a trigger.
The dynode signals do not generate triggers.
The high-gain signals have linear responses for energies up to about 100\,keV
while the dynode signals start to show non-linear responses above about 3,000\,keV.

Signals from the liquid scintillator PMTs are amplified by a gain of 30
and are digitized in a charge-sensitive 62.5\,MSPS ADC (SADC). Unamplified signals from the
muon panels are directly connected to SADCs.
The SADC returns the integrated charge and the time of the input signals.
Triggers from individual channels are generated by field programmable gate
arrays\,(FPGAs) embedded in the FADCs and SADCs.
The final trigger decision
for an event is made by a trigger and clock board\,(TCB)
that also synchronizes the timing of the different channels.

For channels with waveforms that are only non-triggered baselines, the contents are suppressed to zero.
Raw data are converted to a ROOT format~\cite{Brun:1997pa} in the DAQ computer and
automatically transferred daily to CPU farms at the Institute for Basic Science and Yale
University where further processing is performed.
The total trigger rate during a physics run is
28\,Hz, of which crystal FADC triggers constitute 15\,Hz and muon SADC triggers contribute the remaining 13\,Hz.
The total anticipated data size is 7\,TB per year.
In addition, a fraction of the real-time raw-data waveforms
are sent to an online server for monitoring purposes and hour-by-hour monitoring variables are
produced shortly after the raw data creation.

\subsection{DAQ for LS and muon veto}
The LS veto and the muon detector PMTs (60 channels in total) are connected to three SADC modules.
We use a 192\,ns integration time to contain the plastic and liquid
scintillator signals, both of which have decay times less than 10\,ns~\cite{knoll}.

The SADC modules can produce their own triggers and event records.
For this case, the SADC trigger signals are based on the integrated charge.
For the muon counters, 4,000 ADC-count thresholds (integrated charge approximately 763\,pC) are used
to reject PMT noise and $\gamma$-ray induced backgrounds.
Since the observed charges for muon events are typically larger than 12,000 ADC counts,
this trigger-threshold does not reject any muon candidates.
SADC triggers are generated when at least two channels exceed the threshold within a 400\,ns time
window, since any muon that traverses the detector should produce hits in at least two panels.
In addition, a separate trigger is generated if one of the LS PMTs registers a signal\,(with $>$4000
integrated ADC counts) in coincidence with one of the muon panels,
corresponding to events where cosmic-ray muons traverse, or stop inside the LS volume.
There are no triggers that are based only on LS PMT signals. When the SADC modules produce a
trigger, the TCB does not send a trigger signal to the FADC modules; if there is no accompanying
FADC trigger, only data from the three SADC modules are recorded.
Otherwise, the SADC channels provide passive data that are recorded
when FADC triggers are generated by the TCB.

\subsection{Crystal DAQ}
The NaI(Tl) crystal waveforms are digitized by FADCs.
Each FADC module contains four channels and eight FADC modules
are used in COSINE-100.
Their range is from 0 to 2.5\,V with 12-bit resolution.
Similarly to the SADC, the trigger configuration parameters are
uploaded into FPGAs located in each FADC module.

\begin{figure}
  \begin{tabular}{c}
    \includegraphics[width=0.48\textwidth]{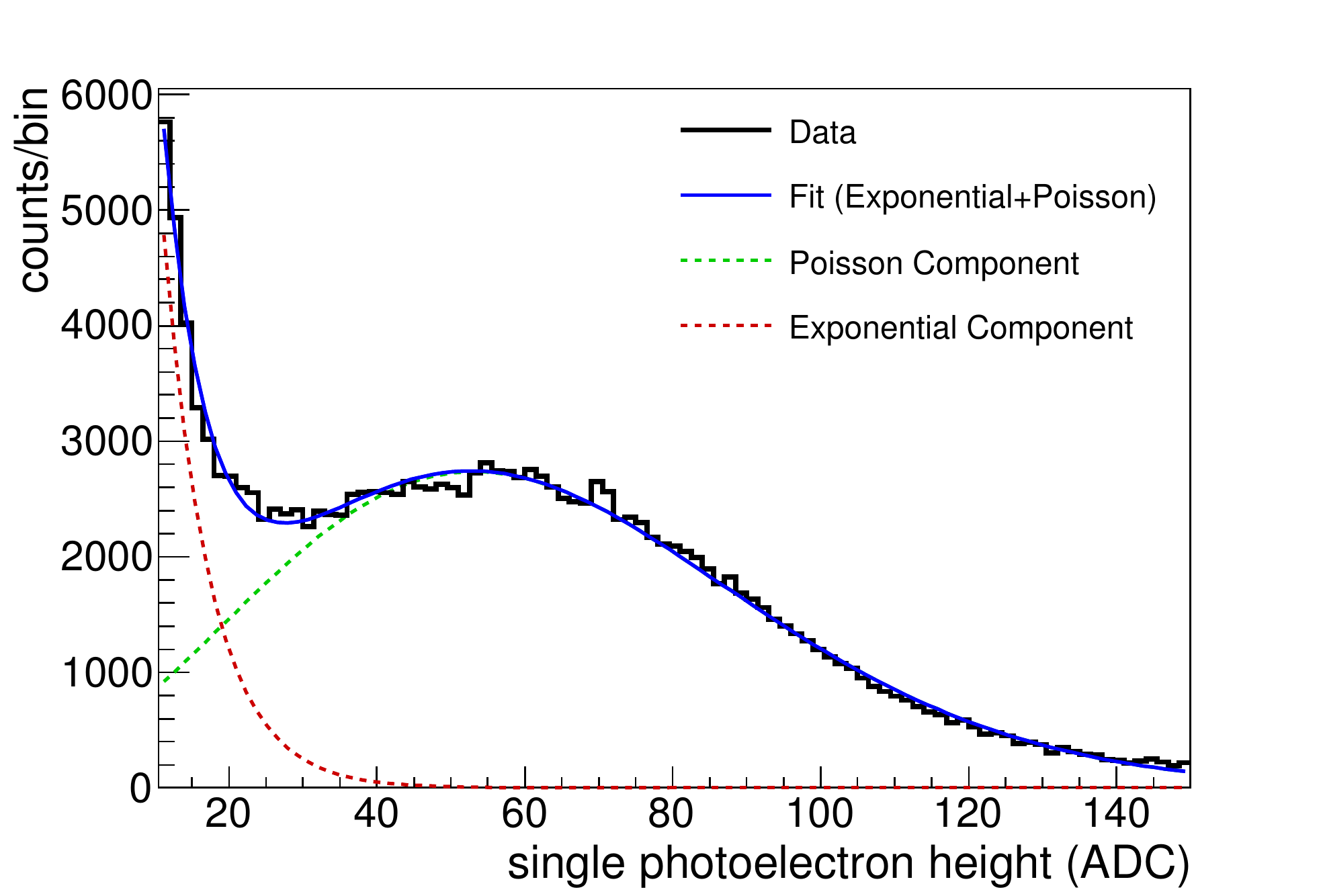}
  \end{tabular}
  \caption{The SPE height spectrum for one of two PMTs in Crystal-1.
  }
  \label{fig:spespec}
\end{figure}

The trigger is generated by high-gain anode signals.
The trigger condition for a NaI(Tl) crystal is satisfied
when the signal crosses a height equivalent to one photoelectron
in both PMTs coupled to a single crystal within a 200\,ns time window.
A ``hit'' is defined as a single photoelectron\,(SPE) with 10 or more ADC counts (greater than 6\,mV)
while a typical SPE signal in the COSINE-100 detector is greater than 25\,ADC counts.
The SPE height spectrum is shown in Fig.~\ref{fig:spespec}.
If at least one crystal satisfies the trigger condition, the TCB transmits trigger signals to all of
the FADCs and SADCs.
All of the crystal, LS, and
muon-detector PMT signal data are stored for FADC triggers.
For each FADC channel, this corresponds to an 8\,$\mu$s-long
waveform starting approximately 2.4\,$\mu$s before the time of the first hit.
For each SADC channel this corresponds to the maximum integrated charge within a 4\,$\mu$s search window
and to the associated time of that maximum.

\section{Slow monitoring system}\label{slowslow}
For stable data-taking and systematic analyses of seasonal variations, it is important to control
and continuously monitor environmental parameters such as detector temperatures, high voltage
variations, humidities, etc.
For this, we employ a variety of sensors for specific monitoring tasks.
These devices are controlled and read out by a common database server and a visualization program.
In this section, we briefly describe the environmental monitoring system for the COSINE-100 experiment. 

We monitor temperatures at various locations using an 8-channel data logger coupled to K-type
thermocouple sensors.  Three of the sensors are placed in contact with the liquid scintillator
inside the copper box.  The others monitor the room and tunnel temperatures. 

\begin{figure*}[htbp]
  \begin{center}
    \includegraphics[width=0.95\textwidth]{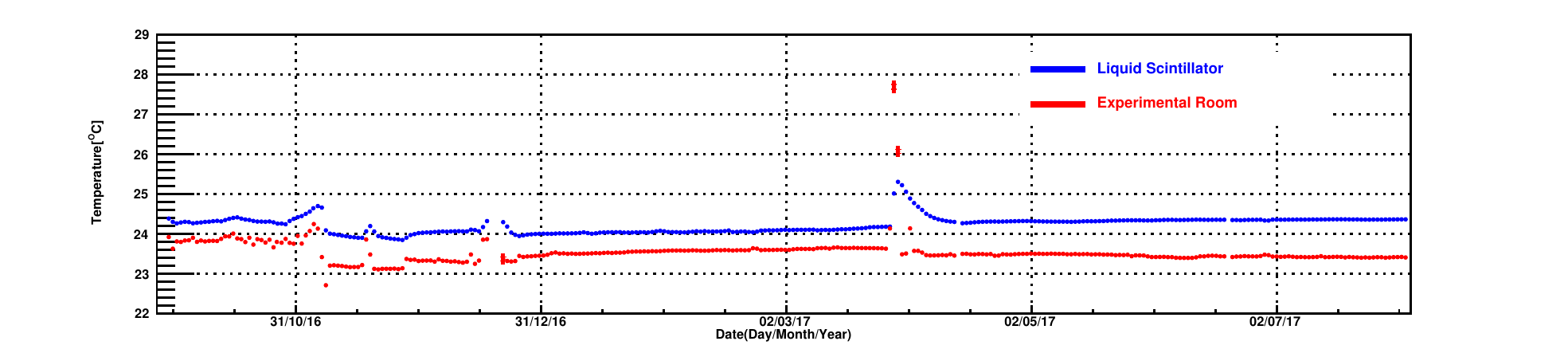}
    \caption{
      Temperature measurements as a function of time in one year period.
      The temperature of the liquid scintillator in the vicinity of Crystal-2
      and one of the detector room temperatures
      are displayed. Note that the offset between the two measurements is due to different sensor calibrations and the high rise in the temperature on March 29, 2017 was due to a sudden air conditioner failure.
    }
    \label{fig:grafana}
  \end{center}
\end{figure*}

High voltages are provided and controlled by a CAEN HV crate that is monitored with software
provided by the company. All of the supplied HVs, currents, and PMT statuses are monitored once
per minute.  We use three analog sensors to measure the relative humidity.
The humidity sensors are connected to the slow-monitoring server via a commercial DAQ module.
The same DAQ module is used to monitor output voltages from the preamplifier system.
The detector-room oxygen level is monitored with a device that has an RS-232 port for serial communication.
As a safety measure the O$_2$-level is prominently displayed in the detector room.
The recorded air-conditioner data include the status of the equipment, the room temperature
and humidity, and various alarms that are transmitted via an RS-485 that is used for continuous
monitoring. All of the monitoring equipment are powered by a 80\,kVA uninterruptible power supply (UPS)
that contains a network-based monitoring module that provides various
protocols. The slow monitoring system checks the UPS status and input/output voltages every 5 seconds.
A commercial RAD7 radon-level monitor samples the air atmosphere in the detector room and records the radon level every 30\,minutes.
We use InfluxDB~\footnote{https://www.influxdata.com} for storing the slow monitoring data and Grafana~\footnote{https://grafana.com}
for visualization of data. Figure~\ref{fig:grafana} shows one of the slow monitoring data panels.

\begin{figure}[!htb]
  \begin{center}
    \includegraphics[width=0.5\textwidth]{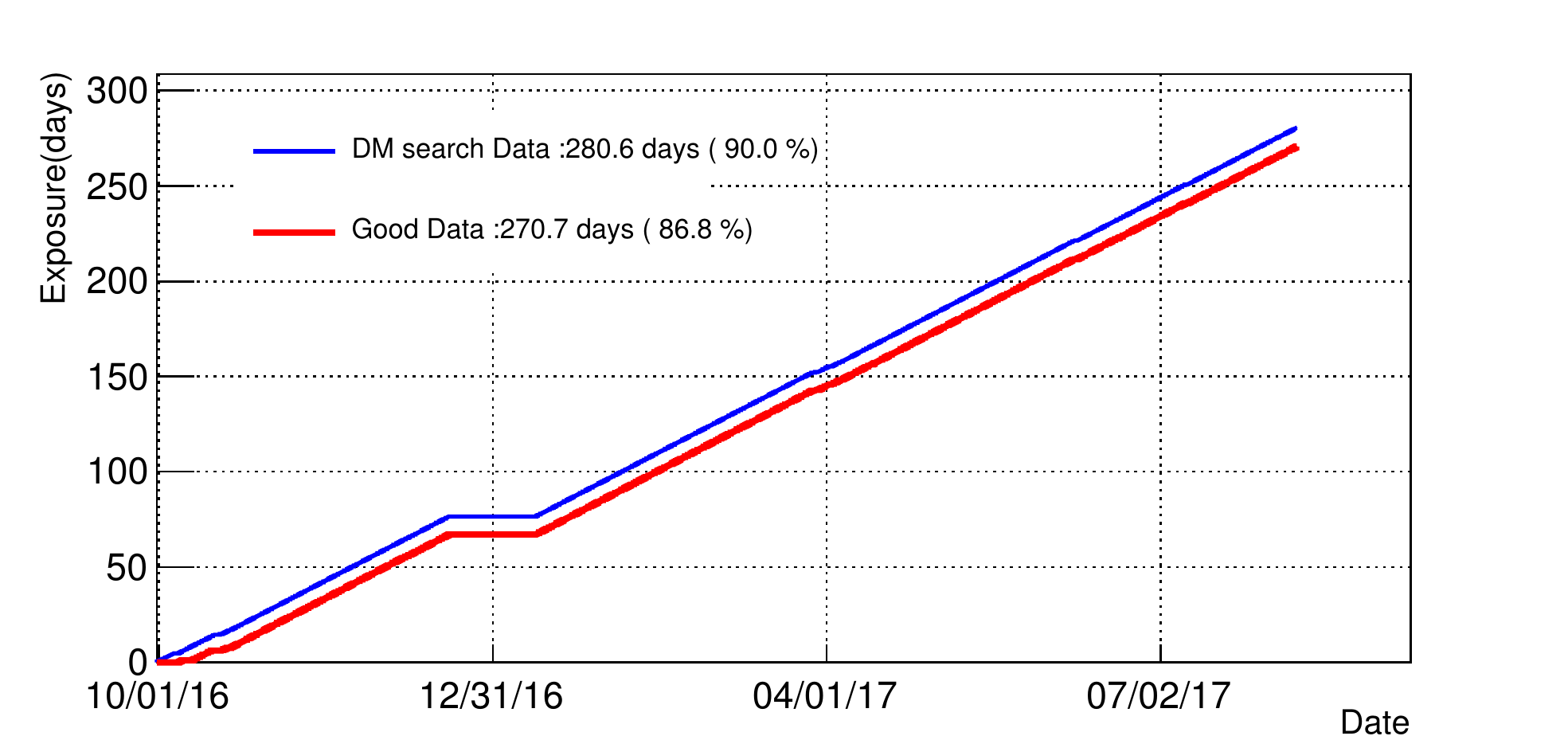}
  \end{center}
  \caption{COSINE-100 exposure versus date.
    The detector stably collects physics data with a 90\,\% live time, of which 96\,\% is
    high-quality. Short instabilities at the beginning of the physics data-taking are mainly due to
    temperature fluctuations in the vicinity of the crystals and DAQ tests.
    The data taking was interrupted on December 20, 2016 for a two-week long calibration campaign
    with external radioactive sources.
  }
  \label{livedays}
\end{figure}
\section{Initial performance of the COSINE-100 detector}\label{datadata}
\begin{figure*}[!htb]
  \begin{center}
    \includegraphics[width=0.95\textwidth]{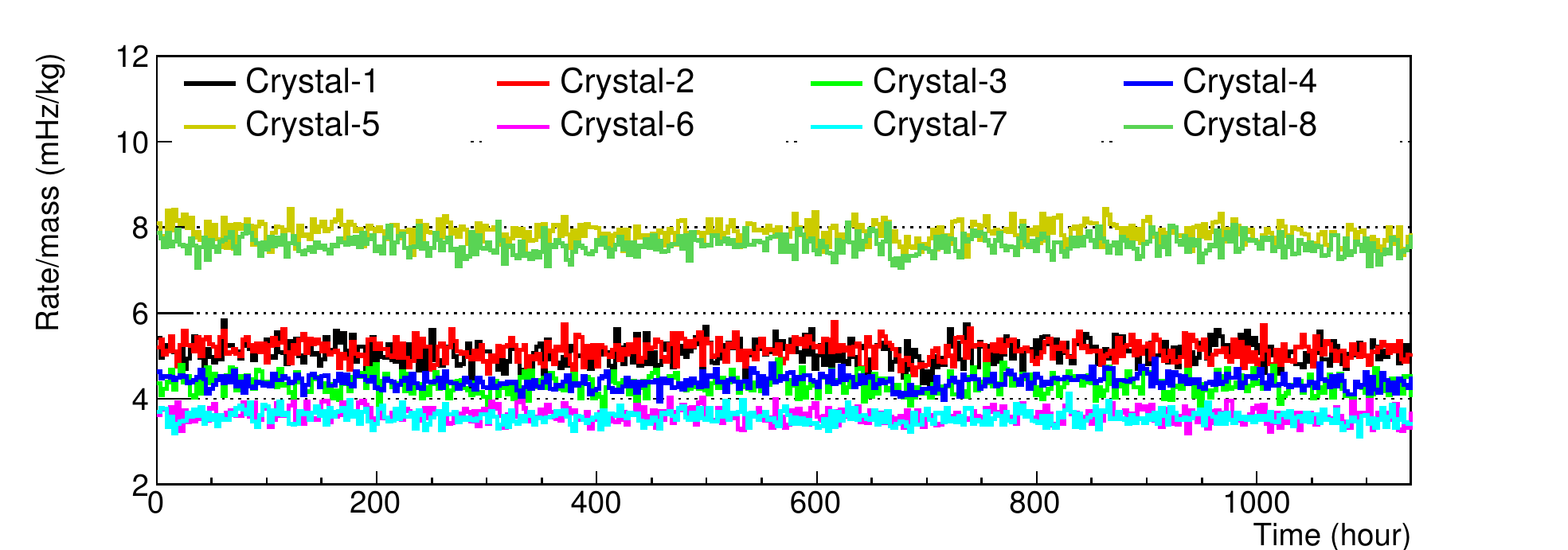}
  \end{center}
  \caption{Crystal trigger rates versus time (in hours) for the first physics run (48 days).
    All of the crystals show stable behavior throughout this running period and the rest of data-taking.
  }
  \label{crystaltriggerrate}
\end{figure*}

After the completion of the detector component installation,
we performed a series of calibrations and began a physics run.
The experiment has been operating stably and collecting physics-quality
data since late September 2016 (see Figs.~\ref{livedays} and~\ref{crystaltriggerrate}).
Monitoring shifts are performed to check data quality in every two hours using automatically generated physics distributions and plots from the slow monitoring system.
Data are defined as a good quality if all DAQ systems are operational with no missing signals,
the LS temperature is not higher than 25$^\circ$C, and the total trigger rate is not higher than 100 Hz for more than 10 minutes.
In this section, we discuss detector calibrations and evaluations of the initial performance.

\subsection{Energy Calibration }
The main goal of the  external radioactive source energy calibration campaign was
to determine the light characteristics of the crystals and the scintillating liquid, 
including light yields, energy scales, and energy resolutions.  The calibration
stability is monitored during data-taking periods by tracking internal $\beta$- and
$\gamma$-ray peaks from radioactive contaminants in the crystals.

At energies above a few hundred keV, calibrations are made using $\gamma$-ray lines
from  $^{137}$Cs and $^{60}$Co sources; $^{137}$Cs produces a mono-energetic $\gamma$-ray
 peak at 662\,keV while $^{60}$Co produces peaks at 1173\,keV and 1332\,keV.  For energies
below 150\,keV, $^{241}$Am and $^{57}$Co radioisotopes with peaks at 59.5\,keV and 122\,keV,
respectively, are used.

\begin{figure*}
  \begin{tabular}{c}
    \includegraphics[width=0.90\textwidth]{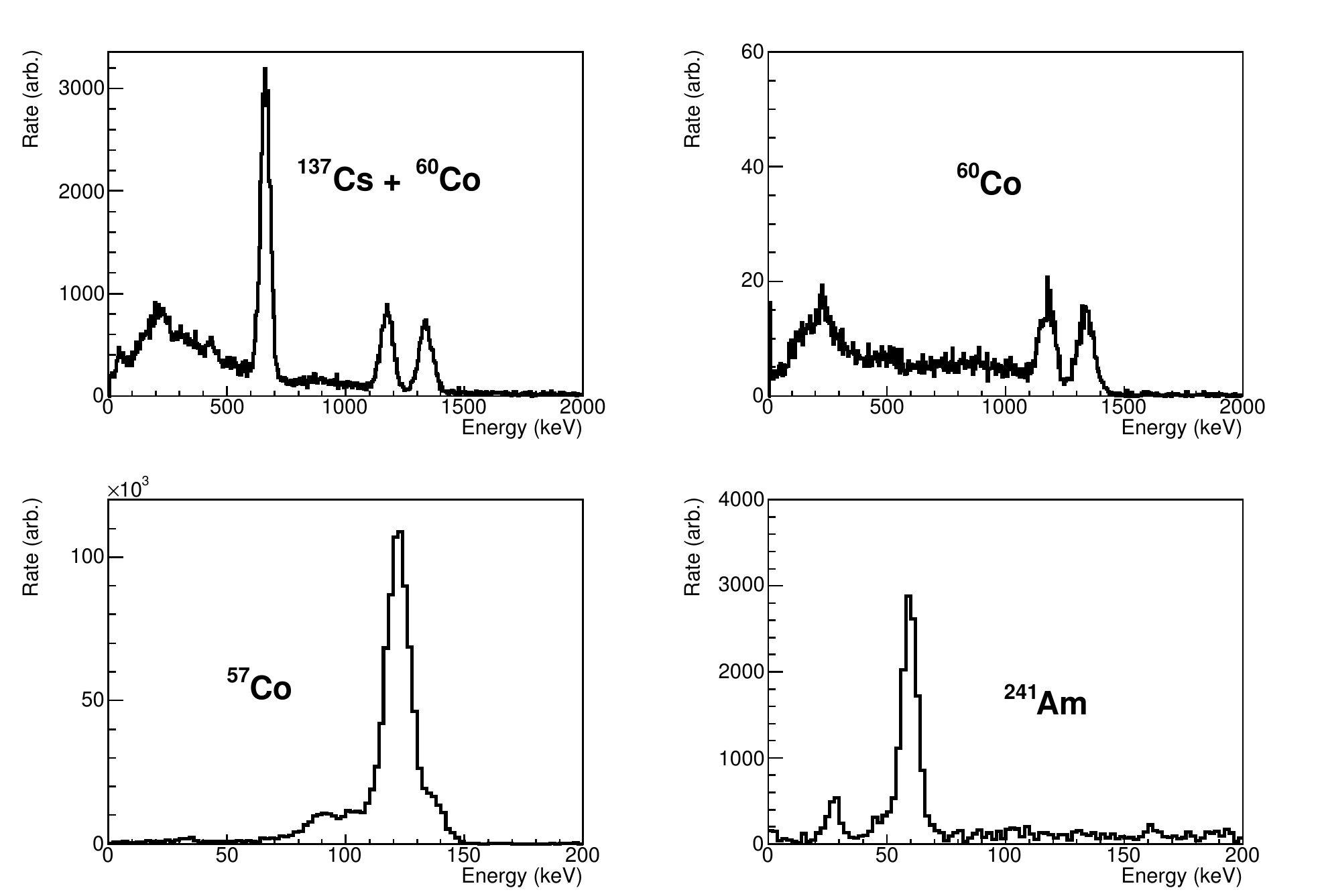}\\
  \end{tabular}
  \caption{The measured calibration energy spectrum in Crystal-6 for
   $^{60}$Co and$^{137}$Cs, $^{60}$Co only, $^{241}$Am and $^{57}$Co calibration sources
   (clockwise from top left). }
  \label{fig:energycalibration}
\end{figure*}
Energy spectra measured with these sources in place are shown in Fig.~\ref{fig:energycalibration}.
The five dominant peaks from the above-mentioned sources, as well as internal radioisotope peaks, are used to set the energy
scale of the crystal spectra based on linear fits.
These energy scales are set separately for the anode readout and the dynode readout.

\begin{figure}
  \begin{tabular}{c}
    \includegraphics[width=0.45\textwidth]{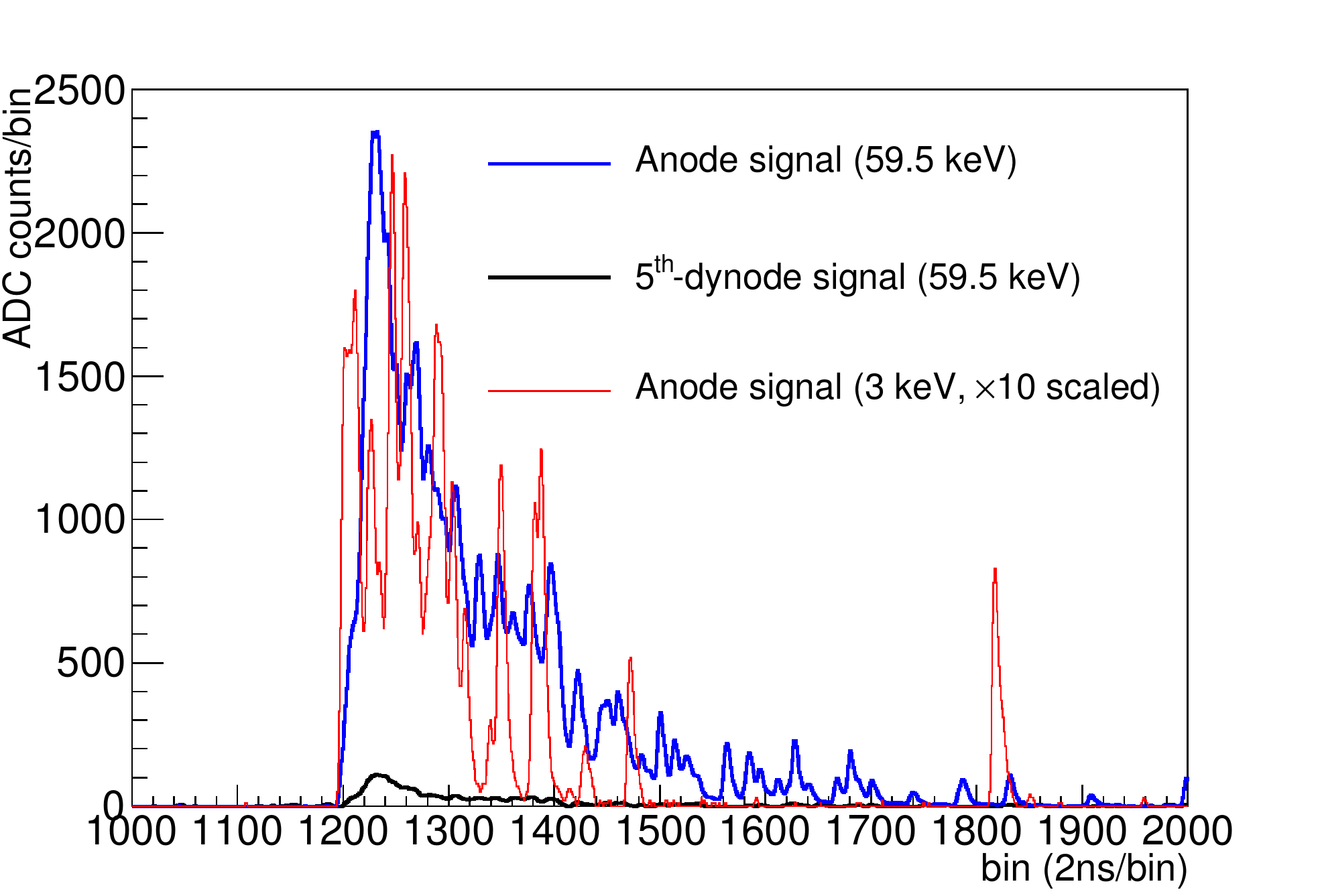}
  \end{tabular}
  \caption{Crystal-1 raw waveforms from a 59.5\,keV \amtwofortyone source and  3\,keV \kforty emission (scaled up
    by a factor of ten) are overlaid for anode signal comparisons.
    For high and low gain comparison, the dynode readout of the same 59.5\,keV signal is also included.
  }
  \label{fig:wave}
\end{figure}
For the low-energy region, the high-gain anode signal waveforms are used.
The energy scale is set by the 
59.5\,keV \amtwofortyone calibration peak. This peak 
is also used to determine the energy resolution and evaluate the crystal light yield.
Additional continuous checks of the energy scale are provided
by the 46.5\,keV \pbtwoten $\gamma$-ray line and the 3\,keV \kforty emission line from internal
contaminants.
From the first three months data analysis,
the \pbtwoten peak positions for the 16 PMTs were measured
to be shifted on average by (0.6$\pm$0.3)\,\% relative to the beginning.
The measured energy resolution, defined as the standard deviation divided by the energy from a Gaussian fit,
is 5\,\% at 59.5\,keV in the Crystal-1 anode readout.
Representative anode waveforms for 59.5\,keV and 3\,keV signals,
and the dynode waveform for the same 59.5\,keV signal, are shown in Fig.~\ref{fig:wave}.

We characterize the PMT responses to SPEs using a sample of isolated
hits occurring in the tails of anode signal waveforms.
This information is used to determine the light yields observed in each PMT
and, from that, each crystal. These spectra and light yields are monitored on a regular
basis in order to monitor the stability level of each PMT.
Figure~\ref{fig:spespec} shows the SPE height spectrum of one of the PMTs in Crystal-1.
A combined fit of a Poisson function to represent the SPE response plus an exponential
to represent a baseline noise component is shown. Results from this fit are used to infer the average number of SPEs in 59.5\,keV \amtwofortyone $\gamma$-ray signals.
With the exception of C5 and C8, the light yields of the COSINE-100 crystals are twice as high as those of the DAMA
crystals~\cite{Bernabei:2008yh}; the light yields are listed in Table ~\ref{activity}.

\subsection{Low energy noise rejection}

Because the hardware trigger threshold is set low, the
DAQ system collects a large number of non-physics events that are primarily caused
by PMT noise that is coincident between the two PMTs coupled to different ends of the same crystal.
These coincident noise events could be due to radioactivity in the PMT glass
and/or circuitry, the discharge of accumulated space charge in one of the PMTs, PMT dark current,
afterpulses of large pulses, etc.  Since PMT-generated noise signals are concentrated
at low energy regions where they could potentially mimic dark matter signals,
these noise events have to be rejected by software selection criteria.

The DAMA group reported signal selection criteria based on time-integrated charge fractions
that efficiently distinguish PMT noise waveforms from those generated by scintillation
events in NaI(Tl) crystals.  The selection exploits the characteristic short decay times
of PMT noise pulses.

\begin{figure}
  \centering
  \includegraphics[width=2.5in]{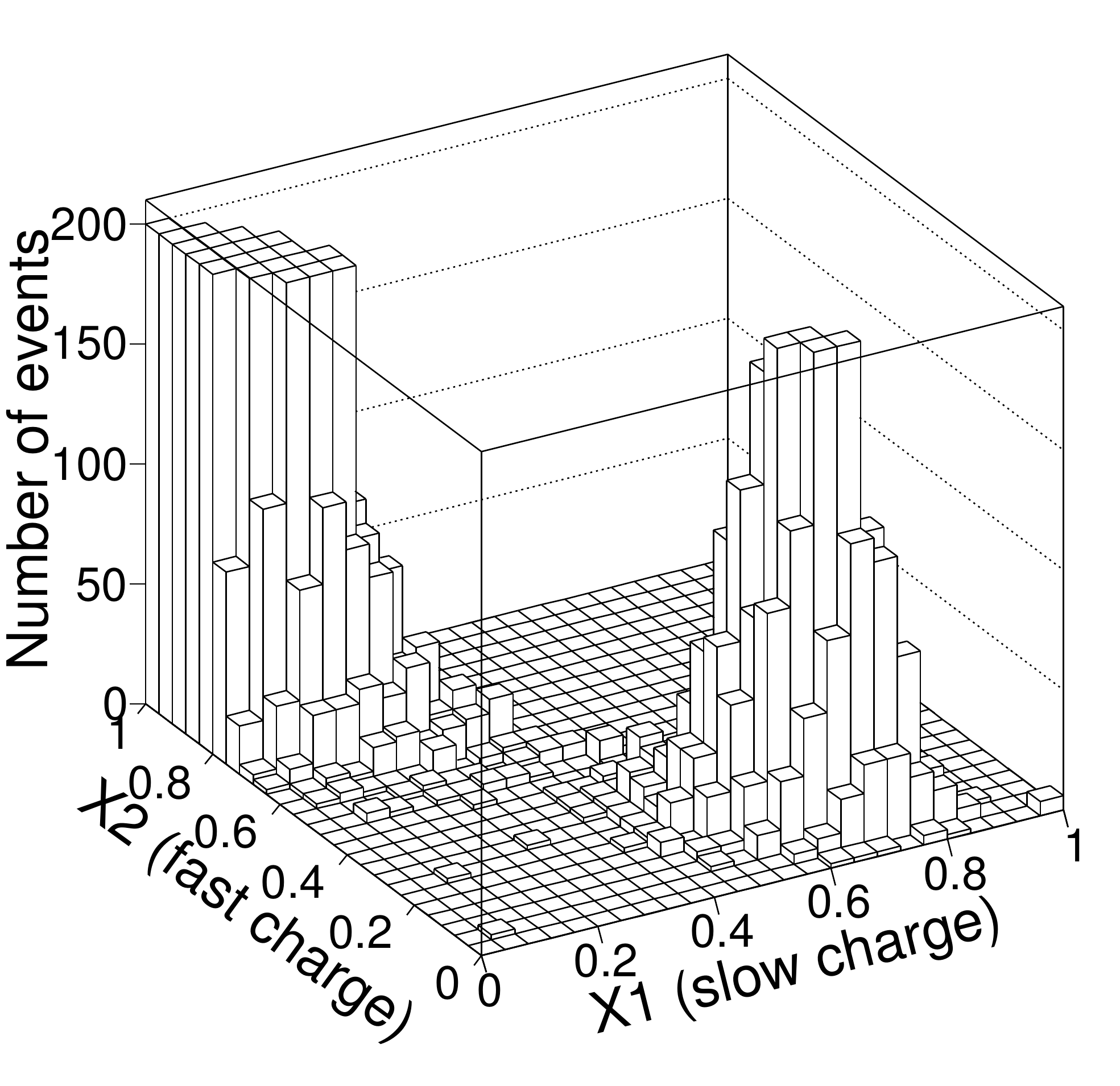}
  \caption{A two-dimensional plot of the fast and slow charge fractions for Crystal-1.
    PMT noise events populate the high X2 and low X1 region while
    signal events have low values of X2 and high values of X1. }
  \label{fig:distribution1}
\end{figure}

DAMA defines the integrated charge in the [0, 50\,ns] time range normalized by the total charge
(integrated over [0, 600\,ns]) as X2 (the fast charge), and that integrated over [100, 600 ns]
time range normalized by the same total charge as X1 (slow charge)~\cite{Bernabei:2008yh,bernabei12had}.
A two-dimensional plot of X1 {\em versus} X2 for low energy COSINE-100 noise and $\beta$-/$\gamma$-ray induced
signal events is shown in Fig.~\ref{fig:distribution1}, where signal events populate the 
high-X1 and low-X2 region while PMT noise events concentrate in the complementary and
well-separated, high-X2 and low-X1 region.
Crystal-specific selection requirements on X1 and X2
typically reject 80\,\% of the PMT-generated noise events above 2\,keV while
retaining 99\,\% of the tagged Compton scattering events at 2\,keV using \cosixty calibration data.

\begin{figure}
  \centering
  \begin{tabular}{c}
    \includegraphics[width=0.45\textwidth]{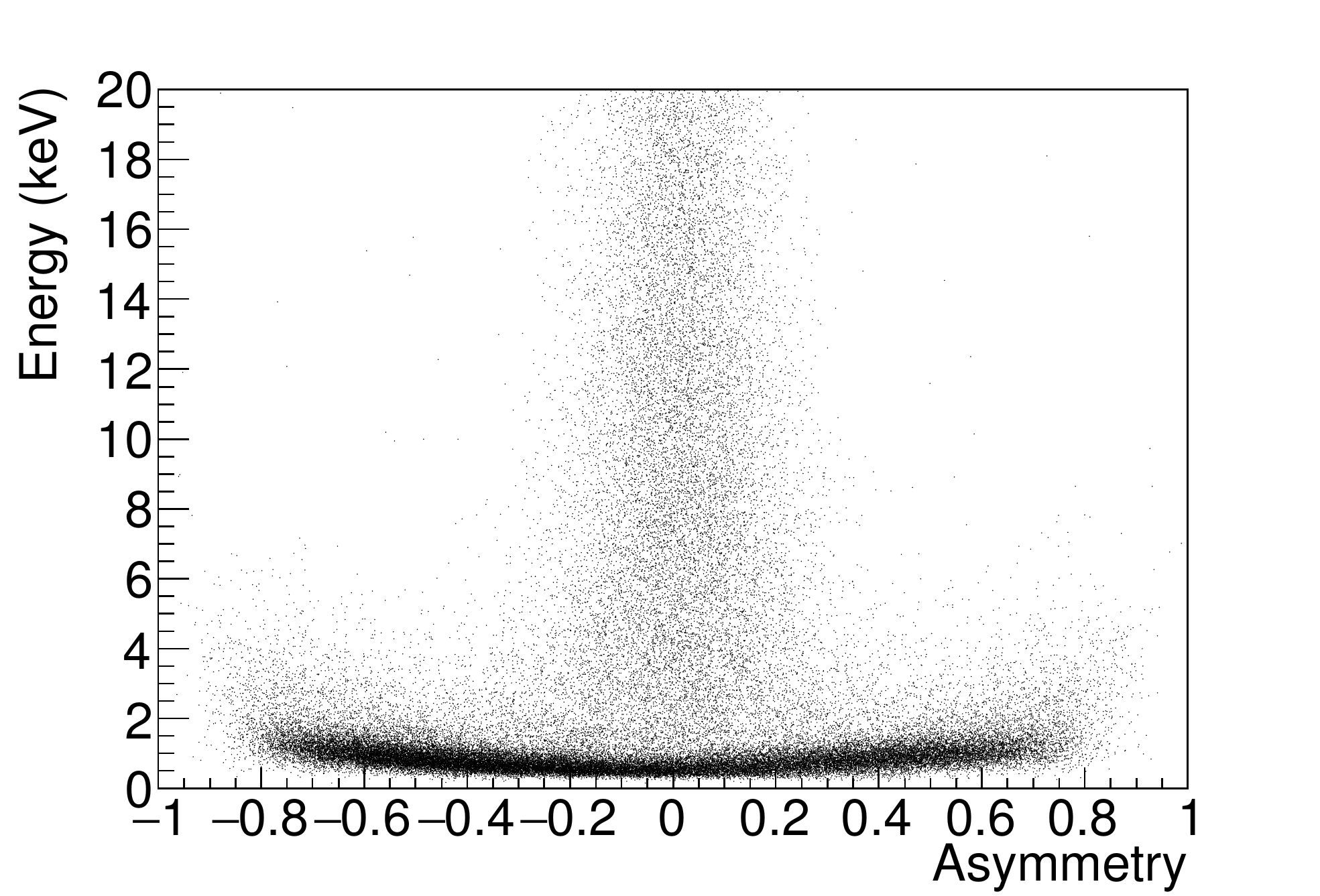} 
  \end{tabular}
  \caption{Energy versus asymmetry for Crystal-1 (see text for the definition of asymmetry). Further noise event reduction is accomplished by a selection criteria
    based on the asymmetry of the total charges measured by the two PMTs. 
    The middle vertical band centered near zero is predominantly due to
    $\beta$-/$\gamma$-ray-induced events inside the crystal, while the highly asymmetric events with
    energies below a few keV are due to events that are likely of PMT origin.
  }
  \label{fig:distribution2}
\end{figure}

Although a large fraction of PMT-noise events are removed by this X1:X2 cut, some
PMT noise-like events still remain.  Therefore, we developed further analysis cuts to remove
these events. One is based on the charge asymmetry between the signal in the two PMTs coupled
to the same crystal, defined as
\[ \rm Asymmetry =  \frac{Q1-Q2}{Q1+Q2}\,, \]
where Q1 and Q2 are the total charges measured by each of the two PMTs.
In the asymmetry distribution shown in Fig.~\ref{fig:distribution2}, the X1:X2 requirement has
already been applied. Many events with apparent energy below 3\,keV have asymmetries that are larger
than those for true signal events that occur near one or the other end of the crystal,
suggesting that these events are caused by visible light produced near one of the PMTs. 
Asymmetric events of this type are only weakly correlated with the X1:X2 requirement,
which exploits the pulse development in time.  To suppress these events, we require that
the absolute value of the asymmetry parameter be smaller than 0.5.
The combined efficiency estimated using the \cosixty events at 2\,keV
with the X1:X2 and asymmetry selections is measured to be 98\,\%,
while rejecting 90\,\% of the PMT noise events.

\begin{figure*}[!htb]
\begin{center}
    \begin{tabular}{cc}
      \includegraphics[width=0.9\textwidth]{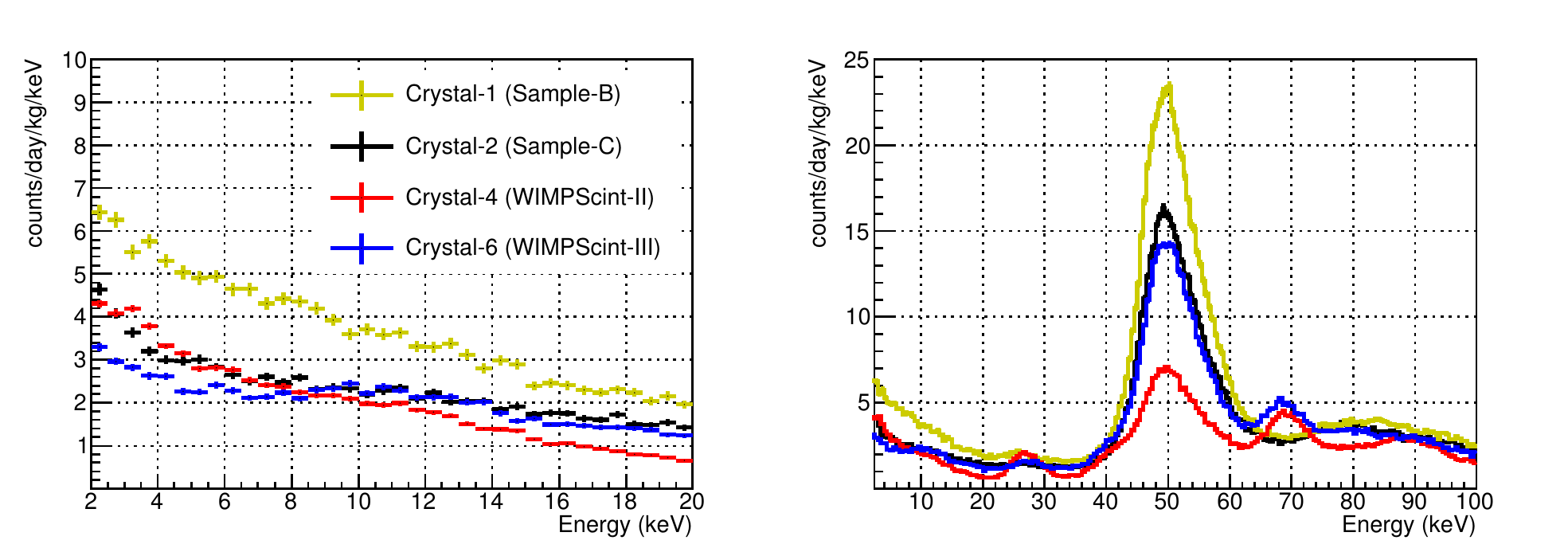}\\
    \end{tabular}
    \caption{ Energy spectrum comparisons for four crystals from different powder samples.
      (left) A zoomed view of the E$\le$20\,keV region of the spectrum.
      The background levels are lowest for Crystal-4 and Crystal-6, which reflect their lower
      \pbtwoten and \kforty contamination levels. The spectra were made after the application
      of the three noise rejection criteria.
      (right) The peak near 50\,keV reflects the \pbtwoten contamination level in each crystal.
      Crystal-4 and Crystal-6
      have been underground for less than one year and so
      cosmogenic peaks (e.g. \ionetwentyfive at 67.8\,keV) are additionally seen.
      These spectra are obtained using 59.5 days of the initial data.
      We have estimated efficiencies at 2\,keV better than 95\% using $^{60}$Co Compton calibration data.
      Therefore, efficiency corrections are not applied.
    }

    \label{spectrum_low}  \end{center}
\end{figure*}

Some noise pulses evade the X1:X2 cut and the asymmetry cut.
These contain merged pulse clusters that correspond
to an anomalously large number of SPEs.  These are removed by placing a limit on the
average number of SPEs per cluster.
For example, we demand that a signal event should contain roughly 2 SPEs or less per cluster at 2\,keV.
This requirement is applied after the X1:X2 and
asymmetry selection conditions and
the total efficiency of all three selections for the \cosixty events at 2\,keV is better than 95\,\%,
while the total PMT noise rejection is 95\,\%.
The low energy spectrum after the application of
the three selection requirements is shown in Fig.~\ref{spectrum_low}.
Studies of
other noise suppression parameters and the development of a multivariate technique
to optimize selections based on these are in progress.

\subsection{$\alpha$ activity and $^{210}$Po  background}

\begin{figure}[!htb]
\begin{center}
\includegraphics[width=0.48\textwidth]{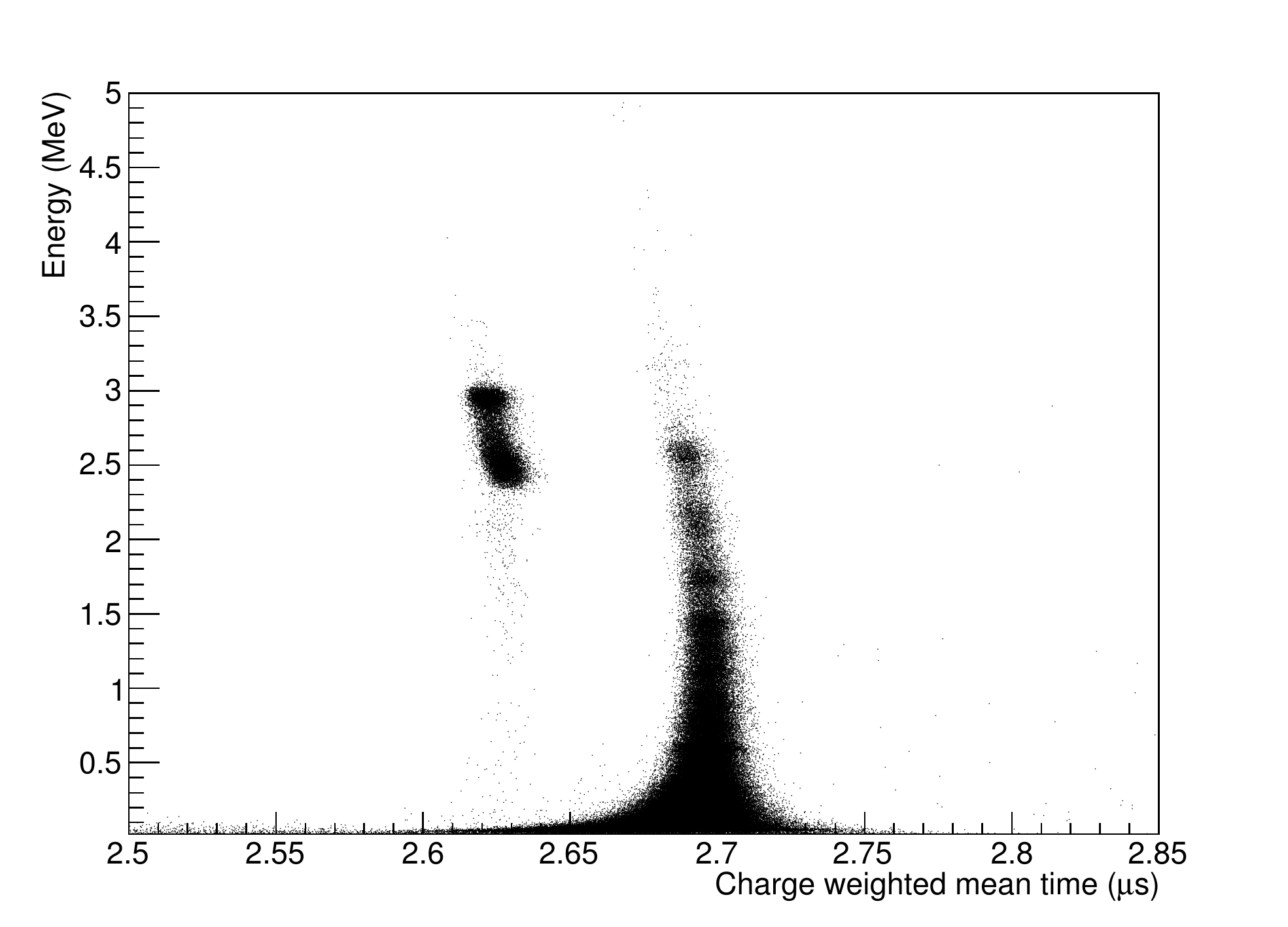}
\end{center}
\caption{Charge-weighted mean time (horizontal) versus energy (vertical) for Crystal-6.
      The island of events with low mean time and high energies is
      due to $\alpha$-induced events and is well separated from $\beta$-/$\gamma$-ray-induced events.
      Due to quenching effects, the measured $\alpha$-energy is lower than its full energy.
      }
  \label{psd}
\end{figure}

Separation of $\alpha$- from $\beta$-/$\gamma$-ray- induced events is achieved by using the
charge-weighted mean-time pulse shape discrimination method~\cite{kykim15}.
The island of events in the Fig.~\ref{psd} scatter plot of charge-weighted mean-time
of a signal waveform {\em versus} its total energy is due to $\alpha$-events, and is well
separated from $\beta$-/$\gamma$-ray-induced events. 

The \utwothirtyeight and \thtwothirtytwo contamination levels measured by $\alpha-\alpha$ and
$\beta-\alpha$ time correlation methods~\cite{kykim15} in the eight crystals
are too low to account for the total observed $\alpha$-rates. This suggests that the bulk of
the $\alpha$-rate is due to decays of \potwoten (E$_\alpha$ = 5.3\,MeV) nuclei that originate from $\beta$-decays of $^{210}$Pb nuclides in the crystals that
occurred sometime during the powder and/or crystal processing
stages~\cite{adhikari16,kykim15}. The $\alpha$-rate for each crystal is listed in Table~\ref{activity}.

\subsection{\kforty  background}
\begin{figure}[!htb]
  \begin{center}
    \begin{tabular}{c}
      \includegraphics[width=0.48\textwidth]{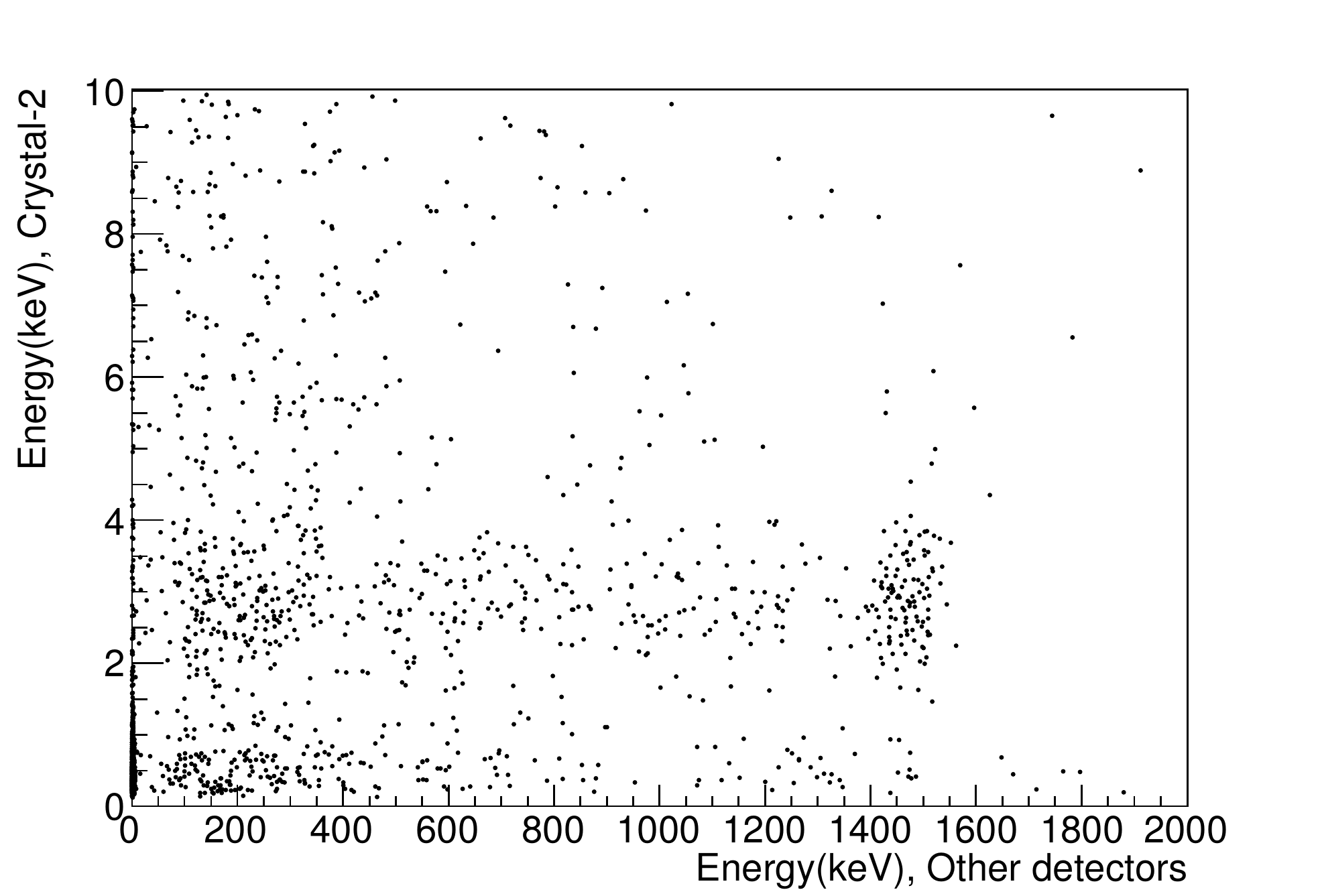}
    \end{tabular}
    \caption{A scatter plot of energy deposition in Crystal-2 (vertical) versus that deposited in other crystals in
      the array (horizontal).
      The \kforty events are identified as the distinct island in the energy spectrum near 3\,keV in Crystal-2
      and 1460\,keV measured in the one of the other crystals.}
  \label{pot}
  \end{center}
\end{figure}

Events generated by decays of \kforty contaminants in the crystals are identified by coincidence
signals between a 3\,keV emission in one NaI(Tl) detector and a 1460\,keV $\gamma$-ray in one of
the other NaI(Tl) crystals or an energy deposition in the LS.
Figure~\ref{pot} shows a scatter plot of the energy in Crystal-2 versus that in other
NaI(Tl) crystals for these multi-hit coincidence events. The \kforty signal forms the island of
events near 3\,keV in the Crystal-2 signal and near 1460\,keV in the other NaI(Tl) crystal.
The \kforty\ background level in each crystal is determined by comparing the measured coincidence
rate with a GEANT4-simulated efficiency using the method described in Ref.~\cite{kykim15}.

The source of \kforty in the NaI(Tl) crystals is almost entirely from contamination that originated in the NaI powder,
and no significant increase in the contamination level is introduced during the crystal growing
procedure~\cite{adhikari16,kykim15}.
The $^{\rm nat}$K measurements for all crystals are listed in Table~\ref{activity},
following the latest analysis with improved statistics.
The $^{\rm nat}$K content in the DAMA crystals is in the 10$-$20\,ppb range~\cite{Bernabei:2008yh},
levels that have been achieved in some of the most recently produced COSINE-100 crystals.

\subsection{Background summary}
\begin{figure}[!htb]
\begin{center}
    \begin{tabular}{lll}
      \includegraphics[width=0.5\textwidth]{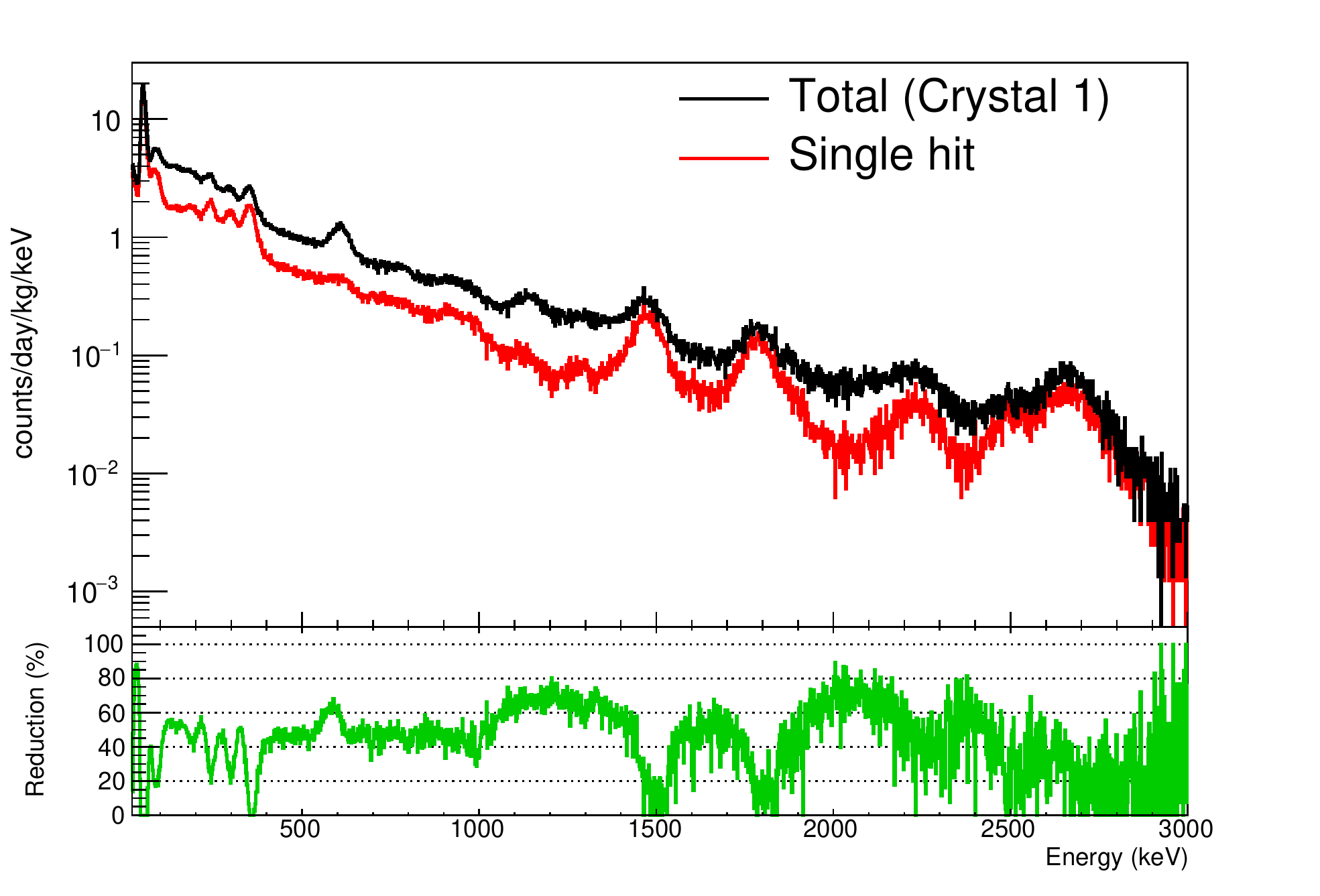} 
    \end{tabular}
    \caption{A comparison of the Crystal-1 energy spectrum before (black line) and after (red line)
      application of the LS veto. The $\gamma$-ray radiation from daughter nuclei of the \utwothirtyeight and \thtwothirtytwo
      decay chains in the external shielding materials are reduced significantly by the LS veto requirement.
      The lower panel shows a reduction in percentage as a function of the energy.
    }
  \label{spectrum_HE}
  \end{center}
\end{figure}

The high-energy $\gamma$-ray spectra show pronounced background peaks: a 1460\,keV line from \kforty
and lines from daughter nuclei in the \utwothirtyeight and \thtwothirtytwo decay chains.
These background levels are reduced by as much as 80\,\% by requiring single-hit crystal events
with no signal in the LS.  Using a GEANT4 simulation, we estimate the efficiency for vetoing
these background components as a function of the crystal's position in the detector array.
Figure~\ref{spectrum_HE} shows a comparison between the high-energy spectrum with and without the LS veto application from real data.

Figure~\ref{spectrum_low} shows a comparison of background levels in
Crystal-1, Crystal-2, Crystal-4 and Crystal-6.
The $^{210}$Pb contamination level was reduced by a factor of four in Crystal-4 as compared to that in Crystal-1,
reflecting improvements in the powder.
Additionally, we are able to achieve a background level close to 2\,counts/day/keV/kg at 6\,keV.
The $^{210}$Pb contamination level in Crystal-6 is lower than in Crystal-1
but higher than that in Crystal-4.
The lower background level in the Crystal-6 energies below 8\,keV and
the flatter distribution between 2--20\,keV compared to Crystal-4 indicate
that this crystal contains less surface $^{210}$Pb contamination relative to that for bulk $^{210}$Pb.
The \kforty level in Crystal-6 is 17 ppb, which is the lowest of the four crystals.
The spectrum of Crystal-7 shows similar characteristics and background levels as those of Crystal-6
since they are created with the same geometry at the same time from the same ingot.
The Crystal-3 spectrum shows a similar shape as Crystal-4 because they are made from the same type of powder.
Crystal-5 and Crystal-8 show about 10\,counts/day/keV/kg at 2\,keV
due to their low light yields and relatively higher background contamination.
The main contributors to the remaining background below 6\,keV are \pbtwoten
$\beta$-decay events, $^{40}$K emissions, and cosmogenic activations of I/Te, $^{22}$Na, $^{109}$Cd, and $^3$H.
The cosmogenically activated
backgrounds will diminish significantly after a few years deep underground at Y2L. 
Analysis tools aimed at lowering the energy threshold are under development.

\subsection{Nuclear recoil calibration}
To characterize the scintillation properties of nuclear recoils,
a study on neutron-induced nuclear recoils is ongoing in a specially prepared neutron calibration facility based on a D--D fusion device that produces 2.42\,MeV mono-energetic neutrons.
Nuclear recoil events are collected for small crystal
targets (2\,cm\,$\times$\,2\,cm\,$\times$\,1.5\,cm) made from each of the crystal ingots placed in the neutron beam.
The neutron scattering angle is used to infer the energy of the
recoil nucleus in the target crystal after selecting on the time-of-flight between the NaI(Tl) crystal signal and the neutron detector signal. 
From this calibration, an event selection efficiency for low energy nuclear recoil events near the threshold region
is determined.

\section{Background model and detector simulations} \label{simu}
\label{section:simulation}
To understand the NaI(Tl) crystal backgrounds and determine WIMP detection
efficiencies for the COSINE-100 experiment,
we have performed Monte-Carlo simulations of the detector and its environment
using the GEANT4 package~\cite{geant4}.
\begin{figure}[!htb]
  \begin{center}
    \begin{tabular}{c}
      \includegraphics[width=0.48\textwidth]{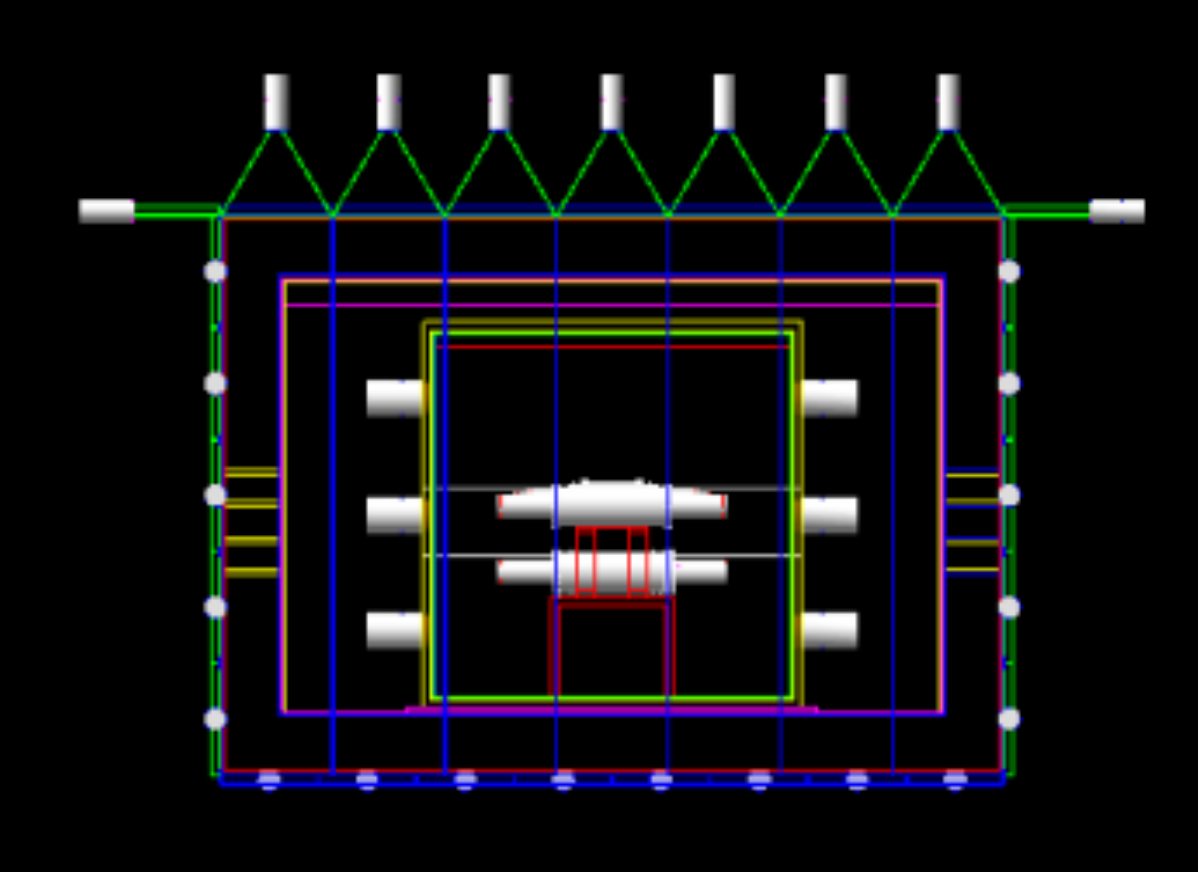} \\
    \end{tabular}
    \caption{A schematic view of the detector geometry.
      The copper encapsulation is shown in white in the middle,
      the acrylic crystal support structure is shown in red,
      the copper box is shown in yellow,
      the lead castle is shown in red, and
      the muon panels are shown in blue.
    }
    \label{detector-setup}
  \end{center}
\end{figure}
The simulation uses the geometry of the COSINE-100 experiment as shown in Fig.~\ref{detector-setup}.

\subsection{Background modeling}
To measure the reduction efficiency of the $^{40}$K 3\,keV background provided by
tagging the accompanying 1460\,keV $\gamma$-ray in one of the other NaI(Tl) crystals or
the LS, and to compare this to the efficiency provided by the other crystals alone,
we generated \kforty decays at random locations inside a NaI(Tl) crystal for the
cases with and without an LS veto.
From these simulations, we determined
the Crystal-6 tagging efficiency by other crystals without LS is 31.7$\pm$0.1\,$\,\%$ and
by the LS only is 64.9$\pm$0.2$\,\%$. The total combined efficiency is 81.7$\pm$0.3\,$\%$.
The efficiency is measured in the crystal energy range between 2 and 6\,keV
by requiring the LS energy deposit be larger than 20\,keV.
Efficiencies vary depending on the crystal location in the detector.
For example, Crystal-1 (at the corner of the array) shows higher coverage by the LS (75\,\%) than
neighboring crystals (17\,\%), but the combined efficiency is similar to that of Crystal-6 (82\,\%).
The tagging efficiency of the 1460\,keV $\gamma$-ray in the LS-only case is lower because
the range of the $\gamma$-ray in the NaI(Tl) crystal is shorter than in the LS.
Therefore, more $\gamma$-rays are stopped in the other crystals than in the LS.
These estimated efficiencies are in agreement with measurements.

Backgrounds from remnants of cosmogenic activation of I and Te radioisotopes still
persist but are declining with lifetimes that are less than 100\,days.
We also observe external backgrounds from the PMTs,
shielding materials, and \rntwotwotwo in the air.
External backgrounds are expected to be mostly tagged
by the LS veto. Thus, the low energy COSINE-100 detector backgrounds are predominantly from
internal sources, especially $^{210}$Pb, of which the main contributions
are from crystal bulk contamination intrinsic to the raw material,
and from \rntwotwotwo exposure during crystal growing and handling procedures.
More detailed information on the background understanding of the NaI(Tl) crystal
can be found in Ref.~\cite{Adhikari:2017gbj}.

\subsection{Detector response and trigger simulation}
To understand the detector response, we are currently developing
a full detector simulation
that allows us to analyze simulated
data using the same framework as real data. The simulation includes photon
generation, photoelectron conversion, amplification, and
FADC digitization. 

The simulation of the scintillation decay time distributions for the NaI(Tl) crystal
signals for electron and nuclear recoils was based on signals generated by a $^{241}$Am source and tuned to the data.
The simulated time distribution of photons detected in the PMTs exhibit a 247\,ns decay time
that is in a good agreement with the known value of 250\,ns~\cite{knoll}.  
We will use digitized simulated waveforms and a trigger logic simulation to
determine the trigger efficiency and, ultimately, perform a detailed simulation of WIMP-induced signals.

\section{Sensitivity of the COSINE-100 experiment} \label{sensi}
The primary goal of \mbox{COSINE-100} is to directly confirm or reject the hypothesis that the annual
modulation observed by DAMA is due to dark matter. Here, we present the projected sensitivity
of the \mbox{COSINE-100} experiment compared with the DAMA-allowed signal regions as interpreted by Ref.~\cite{Savage:2008er}.

We assume the standard halo model of dark matter~\cite{Freese:2012xd} and calculate theoretical modulation
amplitudes of spin-independent WIMP-nucleon interactions in a NaI(Tl) detector as functions of recoil
energy. We further assume an average Earth velocity of 250\,{\rm km/s} and that WIMPs obey a
Maxwellian velocity distribution, with $v_0=220\,{\rm km/s}$, $v_{esc}=650\,{\rm km/s}$, and
$\rho_0=0.3\,{\rm GeV/cm^3}$. To directly compare our results with the modulation amplitudes observed
by DAMA, we assume quenching factors of 0.3 for sodium and 0.09 for iodine, as reported by DAMA~\cite{Bernabei:2008yh}. These theoretical modulation rates are computed for WIMP masses between
1\,and\,10$^3$\,\gev and cross sections between 10$^{-43}$ and 10$^{-37}$\,{\ensuremath{{\rm cm}^2}}.

\begin{figure}[htbp]
	\begin{center}
		\includegraphics[width=0.5\textwidth]{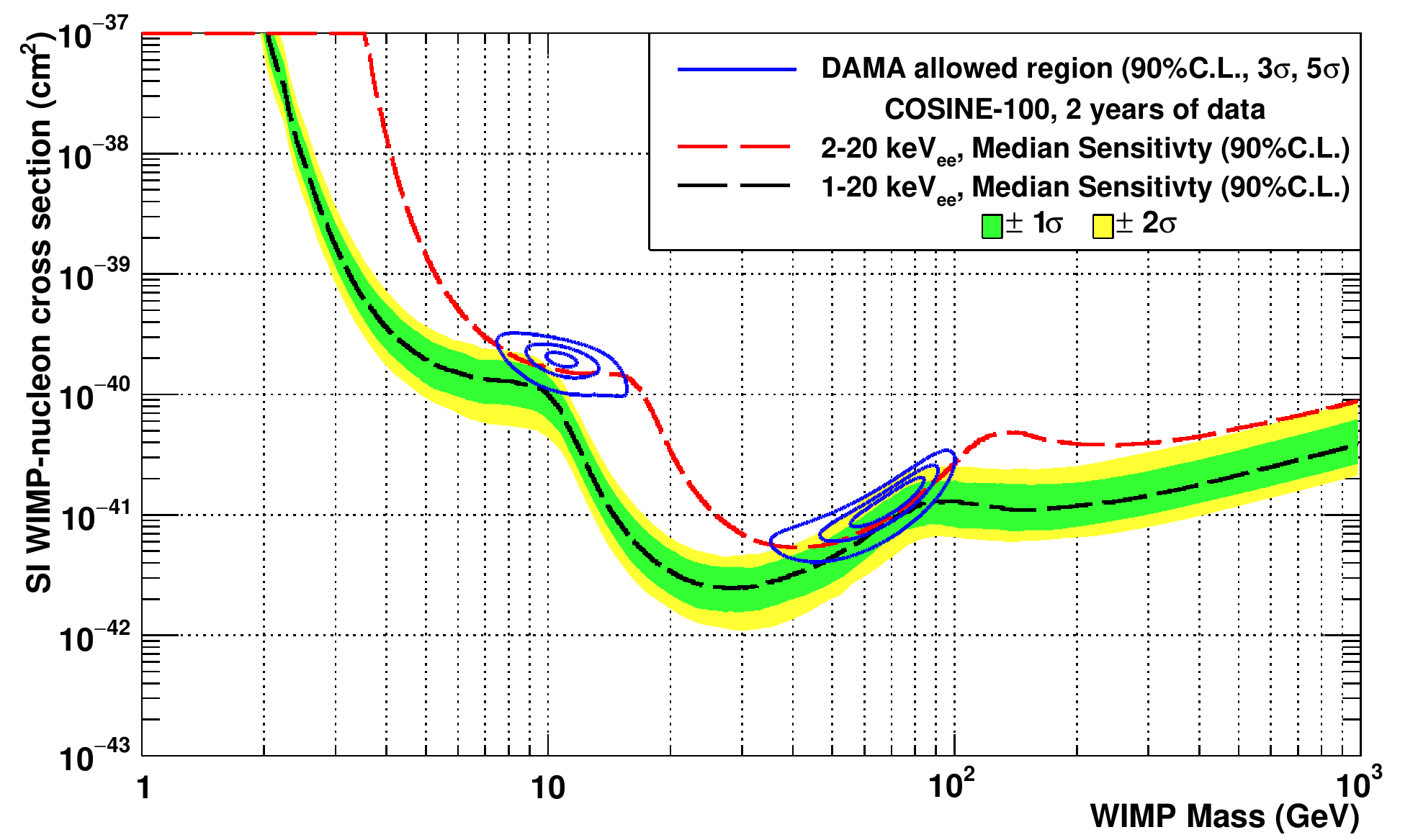}
		\caption{Projected upper limits on the spin-independent WIMP-nucleon cross section using
                 a likelihood analysis. The black curve represents the median exclusion limit of
                 \mbox{COSINE-100} assuming a 1\,keV threshold at a 90\,\%\,C.L. The green and yellow regions
                 represent 1\,$\sigma$ and 2\,$\sigma$ deviations from this median, respectively. The blue
                 contours designate DAMA-allowed regions for spin-independent interactions. A more
                 conservative exclusion limit assuming a 2\,keV threshold for \mbox{COSINE-100} is shown
                 in red.}
		\label{fig:LL}
	\end{center}
\end{figure}

Using this theoretical model, we identify regions in the WIMP phase space compatible with DAMA's observed modulation signal.
The modulation amplitudes obtained from theory are compared with
the modulation amplitude as a function of energy observed by DAMA~\cite{Bernabei:2013xsa}
using a binned likelihood analysis between 1\,and\,20\,keV~\cite{Thompson:2017yvq}.
As can be seen in Fig.~\ref{fig:LL}, two DAMA-allowed regions have been identified,
with the low-mass region corresponding to WIMP-Na scatters and the high-mass region corresponding to
WIMP-I scatters.  Additionally, a $\chi^2$ analysis is performed on these data that gives
results that are consistent with the likelihood analysis.

We also make use of an ensemble of Monte Carlo experiments combined with this theoretical model to establish the
projected sensitivity of the \mbox{COSINE-100} detector in the case of no observed WIMP signal. 
For this analysis, we investigate the sensitivity that will be achieved after two years of data
taking, giving a total exposure of 212\,\kgyr.
A flat representative background of 4.3\,counts/day/kg/keV with no modulation or decay components is assumed based on the currently achieved background levels of the detector.
The flat mean background is calculated by weighting individual crystal masses in the spectrum between 2 and 20 keV.
We simulate the experiment within the background-only
hypothesis by generating histograms of the expected event rate that fluctuate with Poisson statistics
about the 4.3\,counts/day/kg/keV backgrounds for nuclear recoils of various energies. These histograms are binned in
one-day intervals. We then fit a cosine function to the simulated data with a fixed period of one year and a phase of June\,2$^{\rm nd}$.
The fit result is then used to determine the modulation amplitude
observed by \mbox{COSINE-100} for nuclear recoil energies ranging from 1--20\,keV. These simulated
amplitudes are compared to the theoretically predicted modulation amplitudes for various WIMP masses
and cross sections. In total, 200\,iterations of the \mbox{COSINE-100}
experiment are simulated to obtain 90\,\%\,C.L., 3\,$\sigma$, and 5\,$\sigma$ exclusion limits.
The results of the likelihood analysis are shown in Fig.~\ref{fig:LL}.
We also present the projected detector sensitivity for the energy range of 2--20\,keV.

As can be seen in Fig.~\ref{fig:LL}, for the 1\,keV threshold case, the median projected sensitivity of \mbox{COSINE-100} will be
able to exclude the low-mass DAMA-allowed region to a significance of nearly 5\,$\sigma$ at a 90\,\%\,C.L.
and a majority of the high-mass DAMA-allowed region at a 90\,\%\,C.L., assuming no WIMP-nucleon-scattering-like
modulation is observed.
\section{Summary and Conclusions} \label{conc}
The main goal of the COSINE-100 experiment is to independently confirm or dispute DAMA's long-standing annual
modulation signature.
The detector is comprised of eight ultra-low background NaI(Tl) crystals encapsulated in copper
and shielded by several layers of external radioactivity attenuators.
Unlike the DAMA apparatus, the experiment is additionally equipped with cosmic-ray muon panels and a liquid
scintillator veto to tag $^{40}$K-induced events and those that may originate from non-WIMP-induced interactions external to the crystal.
The detector has been taking data since September 30, 2016, and
the fraction of physics-quality data is greater than 95\%.
A variety of control and monitoring systems are in place that collect and record environmental data
that are used in correlation studies with the crystal data.
The initial data performance levels are consistent with expectations and
we expect to continue stable data-taking for the next two years.
With these data, a model-independent analysis will be performed to prove or refute DAMA while
we can also examine a large portion of the WIMP-mass/cross-section parameter regions that are favored by the DAMA results.

\section{Acknowledgments}
We thank the Korea Hydro and Nuclear Power (KHNP) Company for providing underground laboratory space at Yangyang.
This work is supported by:  the Institute for Basic Science (IBS) under project code IBS-R016-A1, Republic of Korea;
UIUC campus research board, the Alfred P. Sloan Foundation Fellowship,
NSF Grants No. PHY-1151795, PHY-1457995, DGE-1122492 and DGE-1256259,
WIPAC,
the Wisconsin Alumni Research Foundation,
Yale University and
DOE/NNSA Grant No. DE-FC52-08NA28752, United States; 
STFC Grant ST/N000277/1 and ST/K001337/1, United Kingdom;
and CNPq and Grant No. 2017/02952-0 FAPESP, Brazil.

\bibliographystyle{plain}

\end{document}